\documentclass[finnish,12pt]{article}
\usepackage[finnish]{thesis}
\pdfoutput=1

\usepackage{graphicx}

\usepackage{amsmath}

\usepackage{url}

\usepackage{amsfonts,amssymb,amsbsy}

\newcommand{\igraph}[2]{
\begin{center}\resizebox*{#1\textwidth}{!}{\includegraphics{#2}}\end{center}
}

\usepackage{comma}

\mathchardef\CommaOrdinary="013B
\mathchardef\CommaPunct   ="613B
\mathcode`,="8000
{\catcode`\,=\active
\gdef ,{\obeyspaces\futurelet\next\CommaCheck}}
\def\CommaCheck{\if\space\next\CommaPunct\else\CommaOrdinary\fi}

\expandafter\let\expandafter\originaleqnarray
  \csname eqnarray\endcsname  
\expandafter\def\csname eqnarray\endcsname
 {\settowidth{\arraycolsep}{$\mskip 0.5\thickmuskip$}\originaleqnarray}
\expandafter\let\expandafter\eqnarraystar
  \csname align\endcsname  
\expandafter\def\csname align\endcsname
 {\settowidth{\arraycolsep}{$\mskip 0.5\thickmuskip$}\eqnarraystar}

\newcommand{\grad}{\textrm{\textdegree}}

\newcommand{\unit}[1]{\relax\ifmmode{\mathrm{#1}}\else{\textrm{#1}}\fi}
\newcommand{\units}[1]{\relax\ifmmode{\ \mathrm{#1}}\else{\ \textrm{#1}}\fi}
\newcommand{\unitf}[2]{\relax\ifmmode{\frac{\mathrm{#1}}{\mathrm{#2}}}\else{$\frac{\mathrm{#1}}{\mathrm{#2}}$}\fi}
\newcommand{\unitfs}[2]{\relax\ifmmode{\ \frac{\mathrm{#1}}{\mathrm{#2}}}\else{$\ \frac{\mathrm{#1}}{\mathrm{#2}}$}\fi}
\newcommand{\D}[1]{\ensuremath{\mathrm{d}#1}}
\newcommand{\vect}[1]{\ensuremath{\mathbf{#1}}}
\newcommand{\trans}[1]{\ensuremath{{#1}^{\mathrm{T}}}}
\newcommand{\Df}[2]{\ensuremath{\frac{\mathrm{d}#1}{\mathrm{d}#2}}}

\newcommand{\I}{\ensuremath{\mathrm{i}}}

\def\us{\units}

\def\condspace{\ensuremath{\ \ \ }}

\begin{document}

\Language{Finnish}{Suomi}

\university{helsinki university of technology}{teknillinen korkeakoulu}

\department{Electronics and Electrical Engineering}%
{Department of Electrical and Telecommunications Engineering}%
{Elektroniikka ja sähkötekniikka}%
{Sähkö- ja tietoliikennetekniikan osasto}%

\degree{BSc}

\author{Juha-Matti Tilli}

\thesistitle{Thesis title}{Puolijohderakenteiden rönt\-gen\-diff\-rakt\-i\-on so\-vi\-tus\-a\-na\-lyy\-si}

\date{12.12.2007}

\supervisor{Prof. Pertti Vainikainen and D.Sc. (Tech.) Juha Mallat}{Prof. Pertti Vainikainen ja TkT Juha Mallat}

\instructor{M.Sc. (Tech.) Aapo Lankinen}{DI Aapo Lankinen}

\makecoverpage

\begin{abstractpage}[finnish]

Työssä käsiteltiin epitaktisista monikerrosrakenteista mitatun
röntgendiffraktiodatan analysointia käyränsovituksella, mitä varten kirjoitettiin tietokoneohjelma. Useasta yhdisteestä
koostuvan kerroksen hilavakio ja Poissonin luku lasketaan Vegardin laista.
Kerrokset ovat jännittyneet pinnan suunnassa, mikä venyttää niitä pinnan
suunnassa ja kohtisuoraan. Kerrokset voivat olla osittain
jännittyneitä.

Mielivaltaisen kiteisen aineen aallonpituusriippuvainen sähköinen
suskeptiivisuus esitetään Fourier-sarjan avulla, jonka komponentit lasketaan
atomien muotokertoimista, anomaalisen dispersion huomioivista Hönlin
korjaustermeistä ja lämpötilariippuvuuden huomioivasta
Debye-Waller-kertoimesta. Epitaktisten monikerrosrakenteiden röntgendiffraktion
intensiteetti tulokulman funktiona lasketaan dynaamiseen diffraktioteoriaan
liittyvän Takagi-Taupin-differentiaaliyhtälön iteraatiokaavamuotoisella ratkaisulla.

Röntgendiffraktiomittauksia analysoidaan sovittamalla simuloitu
diffraktiokäyrä vastaamaan mitattua diffraktiokäyrää. Sovitukseen liittyvä
inversio-ongelma ratkaistiin differentiaalievoluutioalgoritmilla, jossa
käytetään pääkomponenttianalyysiä vähentämään mittausparametrien välisiä
vuorovaikutuksia.

Mittausdatan analysointiin kirjoitettiin Javalla ja Matlabilla
tietokoneohjelma, jonka käyttöliittymä laadittiin käyttäjien palautteen
perusteella. Käyttöliittymällä on hyvin helppo laatia kerrosmalli, jonka
tarkat ominaisuudet saadaan simuloidun käyrän sovituksesta mitattuun
käyrään. Ohjelma osaa käyttää myös GNU Octavea ja siten toimii vaatimatta
mitään kaupallisia osia. 

Diffraktiokäyrien simulointikoodin toimivuutta testattiin vertailemalla sillä
ja kahdella muulla ohjelmalla simuloituja kuvaajia yksi- ja monikerroksisille
InGaAs-, GaAs- ja GaAsN-rakenteille. Ohjelmaa testattiin analysoimalla sillä
mittaussarja, jonka tulokset on julkaistu aiemmin. Tietokoneohjelman
havaittiin antavan samoja tuloksia kuin muutkin ohjelmat, ja sovitusalgoritmin
havaittiin toimivan tehokkaasti.

\end{abstractpage}

\newpage

\mysection{Esipuhe}

Tämä kandidaatintyö on tehty Teknillisen korkeakoulun Mikro- ja nanotekniikan
laboratoriossa. Työ pohjautuu laboratoriossa kahtena viime kesänä tekemääni
kesätyöhön. Työssä käsitellään röntgendiffraktiomittausten teoriaa ja sillä
saadun mittausdatan analysointiin kehittämääni tietokoneohjelmaa. Työ on hieman
normaalia kandidaatintyötä laajempi johtuen vaativasta aiheesta.
Vastaavantyyppinen röntgenheijastusmenetelmä ja siihen liittyvä
tietokoneohjelma jätettiin työn laajuuden vuoksi työn ulkopuolelle.

Haluan kiittää johtaja Veli-Matti Airaksista ja dosentti Markku Sopasta
mahdollisuudesta työskennellä laboratoriossa kahden kesän ajan. Haluan myös
kiittää diplomi-insinööri Aapo Lankista, joka ohjasi kandidaatintyötäni. Ilman
hänen arvokkaita kommenttejaan työ olisi vaikeaselkoisempi ja sisältäisi paljon
virheitä. Lisäksi haluan kiittää diplomi-insinööri Jouni Tiilikaista, joka
tutustutti minut puolijohteisiin, röntgenheijastukseen ja geneettisiin
optimointialgoritmeihin, mitä ilman en olisi tehnyt tätä työtä.

\vspace{5cm}
Otaniemi, 12.12.2007

\vspace{5mm}
{\hfill Juha-Matti Tilli \hspace{1cm}}

\newpage

\addcontentsline{toc}{section}{Sisällysluettelo}
\tableofcontents

\mysection{Symbolit ja lyhenteet}

\subsection*{Symbolit}

\def\popvect{\ensuremath{\vect{u_\mathnormal{i}}}}
\def\newvect{\ensuremath{\vect{u_\mathnormal{i}}}'}
\def\bestvect{\ensuremath{\vect{u_\mathrm{b}}}}
\def\beforemut{\ensuremath{\vect{u_\mathnormal{i,\mathrm{mut}}}}}
\def\aftermut{\ensuremath{\vect{u_\mathnormal{i,\mathrm{mut}}}'}}
\def\mutvect{\ensuremath{\vect{w_\mathnormal{i}}}}
\def\randfirst{\ensuremath{\vect{v_\mathnormal{i,\mathrm{r}_1}}}}
\def\randsecond{\ensuremath{\vect{v_\mathnormal{i,\mathrm{r}_2}}}}
\def\randthird{\ensuremath{\vect{v_\mathnormal{i,\mathrm{r}_3}}}}
\def\randk{\ensuremath{\vect{v_\mathnormal{i,\mathrm{r}_k}}}}

\begin{tabular}{ll}
$\vect{a_1}$, $\vect{a_2}$, $\vect{a_3}$ & Hilavektorit \\
$\vect{b_1}$, $\vect{b_2}$, $\vect{b_3}$ & Käänteishilavektorit \\
$\vect{r}$ & Paikkavektori \\
$\vect{h}$ & Käänteishilan vektori, eli $h\vect{b_1} + k\vect{b_2} + l\vect{b_3}$, jossa $h, k, l \in \mathbb{Z}$ \\
$\vect{E}$, $\vect{H}$, $\vect{D}$, $\vect{B}$ & Sähkö- ja magneettikentän voimakkuudet ja vuontiheydet \\
$\vect{P}$ & Polarisaatio eli dipolimomentti yksikkötilavuutta kohti \\
$\chi(\vect{r})$, $\chi_m(\vect{r})$ & Sähköinen ja magneettinen suskeptiivisuus paikan funktiona \\
$\chi_{\vect{h}}$, $\chi_{\vect{\overline{h}}}$ & Vektoreita $\vect{h}$ ja $-\vect{h}$ vastaavat suskeptiivisuuden Fourier-kertoimet \\
$\I$ & Imaginääriyksikkö \\
$V$ & Yksikkökopin tilavuus \\
$\nu$ & Poissonin luku \\
$R_x$, $\sigma_{xx}$, $\epsilon_{xx}$ & Relaksaatio, jännitys ja venymä suunnassa $x$ \\
$d_{x}$, $d$ & Kiteen hilavakio suunnassa $x$, diffraktiotasojen välinen etäisyys \\
$\omega$, $\omega_0$ & Kulmataajuus, ominaiskulmataajuus \\
$\lambda$ & Aallonpituus \\
$\vect{k}$ & Aaltovektori, jonka suuruus $|\vect{k}| = \frac{2\pi}{\lambda}$ \\
$m_e$ & Elektronin massa \\
$Z$ & Atomiluku eli elektronien määrä atomissa \\
$\vect{x}$ & Elektronipilven keskimääräisen paikan poikkeama tasapainosta \\
$\rho_e(\vect{r})$ & Elektronitiheys paikan funktiona \\
$F = f_1 + \I f_2$ & Atomin sirontakerroin ja sen reaali- ja imaginääriosat \\
$f$, $f'$, $f''$ & Atomin muotokerroin ja Hönlin korjaustermit \\
$B$ & Debye-Waller-B-kerroin \\
$s$ & Muotokertoimen laskennassa käytetty suure, määritelmä $s = \frac{|\vect{h}|}{4\pi}$ \\
$p$ & Okkupaatioluku \\
$\theta_B$ & Braggin kulma \\
$\theta$ & Tulevien ja lähtevien säteiden kulma heijastavista tasoista \\
$\phi$ & Heijastavien tasojen ja pinnan välinen kulma \\
$e$ & Alkeisvaraus (luonnonvakio) \\
$c$ & Valon nopeus (luonnonvakio) \\
$\epsilon_0$, $\mu_0$ & Tyhjiön permittiivisyys ja permeabiliteetti (luonnonvakioita) \\
$r_e$ & Klassinen elektronin säde (luonnonvakio) \\
$\vect{C}$ & Kovarianssimatriisi \\
$\vect{P}$ & Populaatiomatriisi \\
$\vect{T}$ & Koordinaattirotaatiomatriisi \\
$\popvect$, $\bestvect$ & Populaation $i$. yksilö ja paras yksilö \\
$\newvect$ & Uusi $i$. yksilö jota verrataan populaation $i$. yksilöön \\
$\beforemut$ & Mutatoitava yksilö \\
$\aftermut$ & Mutatoitu yksilö \\
$\mutvect$ & Mutaatiovektori \\
$\randk$ & Populaatiosta satunnaisesti valittuja yksilöitä, $k \in \{1,2,3\}$ \\
\end{tabular}

\subsection*{Merkinnät}

\begin{tabular}{ll}
$x$ & Skalaari \\
$\vect{v}$ & (Pysty)vektori \\
$\vect{\hat{v}}$ & Yksikkövektori \\
$\vect{A}$ & Matriisi (poikkeus: $\vect{E}$, $\vect{D}$, $\vect{B}$, $\vect{H}$ ja $\vect{P}$ vektoreita) \\
$\vect{u}\cdot\vect{v}$ & Kahden vektorin skalaaritulo \\
$(hkl)$ & Millerin indeksi tasolle, jonka normaalivektori on $h\vect{b_1} + k\vect{b_2} + l\vect{b_3}$ \\
$(\overline{h}k\overline{l})$ & Millerin indeksi, jossa osa indekseistä on negatiivisia \\
$\trans{\vect{A}}$ & Matriisin transpoosi \\
$\vect{A}^{-1}$ & Käänteismatriisi \\
$\left[x y z\right]$ & $1 \times 3$-vaakavektori \\
$\left[\vect{v_1} \vect{v_2} \vect{v_3}\right]$ & $3 \times 3$-matriisi, jonka sarakkeita ovat pystyvektorit $\vect{v_1}$, $\vect{v_2}$ ja $\vect{v_3}$ \\
\end{tabular}

\subsection*{Lyhenteet}

\begin{tabular}{ll}
XRD        & röntgendiffraktio \\
DE         & differentiaalievoluutio \\
PCA        & pääkomponenttianalyysi \\
ICA        & riippumattomien komponenttien analyysi \\
SI         & kansainvälinen yksikköjärjestelmä \\
CGS        & senttimetri-gramma-sekunti-järjestelmä \\
\end{tabular}

\clearpage
\storeinipagenumber
\pagenumbering{arabic}
\setcounter{page}{1}

\section{Johdanto}
\thispagestyle{empty}

Röntgensäteilyn käyttökelpoisuus aineen rakenteen tutkimisessa perustuu
röntgensäteilyn aallonpituuteen, joka on suuruusluokkaa $0{,}1\ \textrm{nm}$.
Atomien välinen etäisyys kiinteässä aineessa on myös samaa suuruusluokkaa.
Tästä syystä röntgensäteily on herkkä ilmiöille atomien etäisyyden
mittakaavassa. \cite[s.~11]{XSS}

Ainetta, jossa atomit ovat järjestäytyneet säännöllisesti, kutsutaan
kiteiseksi. Atomien elektronipilvi sirottaa säteilyä heikosti joka suuntaan.
Koska atomit ovat järjestäytyneet säännöllisesti kiteessä, kaikista eri
atomeista tiettyihin suuntiin sironneen säteilyn kulkeman matkan ero on
aallonpituuden moninkerta. Sironnut säteily interferoi vahvistavasti. Näissä
suunnissa havaitaan voimakasta röntgensäteilyä. Ilmiötä kutsutaan
diffraktioksi. Diffraktio on puhtaasti aaltoliikkeeseen liittyvä ilmiö, joten
se havaitaan myös muilla sähkömagneettisen säteilyn aallonpituuksilla, jos
sopiva diffraktiohila on käytettävissä. Vaikka röntgensäteilyn vuorovaikutus
aineen kanssa on heikkoa, atomeita on kiinteässä aineessa niin tiheästi, että
diffraktioilmiö pystytään havaitsemaan. Röntgendiffraktiolla voidaan tutkia
mitä tahansa kiderakenteista ainetta. Tärkein selville saatava aineen
ominaisuus on kiderakenne, eli atomien paikat ja niiden väliset etäisyydet
säännöllisessä hilassa. Jos diffraktio mitataan hyvin ohuesta kiteestä
röntgensäteilyn tulokulman funktiona, suurimman intensiteetin kulman eli
diffraktiokulman ympärillä havaitaan intensiteetin värähtelyä kulman funktiona,
josta voidaan määrittää kiteen paksuus.

Jos näytteen rakenne tunnetaan täydellisesti, diffraktioilmiötä 
voidaan käsitellä laskennallisesti klassisen sähkömagnetismin teorialla.
Käsittely perustuu Maxwellin lakeihin, joissa otetaan huomioon röntgensäteilyn
ja aineen vuorovaikutusta kuvaava sähköinen suskeptiivisuus. Näin saadaan
laskettua teoreettisesti, minkälaisia mittaustulosten pitäisi olla. Yhtälöt
ovat kuitenkin niin monimutkaisia, että mittaustuloksista ei voida laskea
yksinkertaisten toimenpiteiden avulla näytteen rakennetta. Tämän tyyppisiä
ongelmia kutsutaan käänteisongelmiksi. Erilaisia rakennevaihtoehtoja
kokeilemalla voidaan yrittää etsiä sellainen rakenne, jonka mittaustulokset
vastaavat mahdollisimman hyvin laskennallista käsittelyä. Jos näytteen rakenne
on monimutkainen, kokeiltavien vaihtoehtojen määrä on valtava. Nykyiset
tietokoneet eivät pysty käymään kaikkia vaihtoehtoja läpi kohtuullisessa
ajassa. Kokeiltavien vaihtoehtojen määrää on siis rajoitettava jollain
älykkäällä algoritmilla, jotta rakenne saataisiin selvitettyä
mittaustuloksista riittävän nopeasti.

Tämän työn tavoitteena on kirjoittaa tietokoneohjelma, joka pystyy 
laskemaan fysikaalisen mallin avulla röntgendiffraktion 
intensiteetit monikerroksisista puolijohderakenteista ja selvittämään
mittausdatan perusteella rakenteiden ominaisuuksia etsimällä käänteisongelmalle
numeerisesti ratkaisu. Tietokoneohjelmiin tehdään tehokkaat käyttöliittymät,
jotta käyttäjä pystyisi helposti rajaamaan mitattavat suureet sellaisille
väleille, että käänteisongelman ratkaisualgoritmi löytää tarkat mitattavien
suureiden arvot tehokkaasti. Ratkaisualgoritmin on tarkoitus olla niin tehokas,
että se pystyy löytämään käänteisongelman ratkaisun monimutkaisillekin
rakenteille.

Kaikissa yhtälöissä käytetään yksikköjärjestelmänä SI-järjestelmää.
Käänteishila ja kompleksinen sähköinen suskeptiivisuus on mahdollista
määritellä usealla toisistaan poikkeavalla tavalla. Tässä työssä käänteishila määritellään siten, että
$\exp(\vect{h}\cdot\vect{k}) = 1$, ja kompleksinen sähköinen suskeptiivisuus
lasketaan aikariippuvuustekijällä $\exp(\I\omega t)$, mikä tekee
suskeptiivisuuden imaginääriosasta negatiivisen.

Luvussa \ref{crystals} esitetään kiteisten aineiden rakenteisiin liittyviä
peruskäsitteitä. Luvussa esitellään kiteiden rakenteeseen olennaisesti liittyvä
hila, joka on vektoriavaruuden kaltainen matemaattinen käsite. Myös
röntgendiffraktioilmiön kannalta olennainen käänteishila esitellään.
Luvussa \ref{epitaxy} käsitellään työn kannalta olennaista puolijohdekerrosten
kasvatusmenetelmää, epitaksiaa. Luvun pääpaino on kerrosten elastisuus.

Luvussa \ref{susc} kerrotaan, miten röntgensäteilyn ja aineen välistä
vuorovaikutusta kuvaava suskeptiivisuus lasketaan erilaisille aineille. Tämä
luku on työn teoreettisin, ja se on vaikea ymmärtää asiaan perehtymättömille.
Suskeptiivisuuden laskenta vaatii useasta lähteestä saatavan tiedon yhdistelyä,
joten suskeptiivisuuden laskenta esitetään yksityiskohtaisesti.

Luvussa \ref{xrd} käsitellään varsinaista työhön liittyvää
mittausmenetelmää, eli röntgendiffraktiota.
Röntgendiffraktioon liittyvä fysiikka esitetään vain
lyhyesti, koska laskentaan liittyvä fysiikka perustuu vain
yhteen lähteeseen. Asiasta kiinnostunut lukija saa lähteestä \cite{bartels} lisätietoa.

Mittaustulosten analysointiin liittyvästä käänteisongelmasta kerrotaan luvussa
\ref{inverse}. Luvussa \ref{DE} esitellään työssä käytettävä algoritmi, jolla
käänteisongelma ratkaistaan. Käänteisongelma ratkaistaan tietokoneohjelmalla,
jonka rakennetta, toimintaa ja käyttöliittymää kuvataan luvussa \ref{software}.

Luvussa \ref{gidsl} verrataan tässä työssä kirjoitetulla
röntgendiffraktio-ohjelmalla ja kahdella olemassaolevalla ohjelmalla,
GID\_sl:llä ja Philipsin röntgendiffraktometrin omalla ohjelmalla, simuloituja
käyriä. Luvussa \ref{meas} esitellään muutaman yksinkertaisen mittauksen
analysointi ja vertaillaan mittaustuloksia aiemmin mitattuihin tuloksiin. Lisäksi
tässä luvussa testataan, toimiiko sovitusalgoritmi monimutkaisilla kerrosrakenteilla.

\clearpage
\section{Teoreettinen tausta}

\subsection{Kide ja hila \label{crystals}}

Kiteisessä aineessa atomit ovat asettuneet säännölliseen rakenteeseen, joka
toistuu kolmessa ulottuvuudessa täyttäen koko avaruuden. Rakenteen toistuva
yksikkö on suuntaissärmiö, jota kutsutaan yksikkökopiksi. Yksikkökopin
määrittelee kolme vektoria, joita kutsutaan hilavektoreiksi ja jotka merkitään
$\vect{a_1}$, $\vect{a_2}$ ja $\vect{a_3}$. Kun tiedetään
hilavektorit ja atomien paikat yksikkökopin sisällä, kidehila on
yksikäsitteisesti määritelty. Jos atomi sijaitsee yksikkökopin sisällä paikassa
$\vect{r}$, kaikki vastaavien atomien paikat koko hilassa $\vect{r'}$
saadaan yhtälöstä (\ref{atompos}). \cite{cryst}

\begin{equation}
\vect{r'} = \vect{r} + h\vect{a_1} + k\vect{a_2} + l\vect{a_3}, \condspace h, k, l \in \mathbb{Z} \label{atompos}
\end{equation}

Yksikkökoppi ei tietenkään ole yksikäsitteinen, koska esimerkiksi kahdesta
vierekkäisestä yksikkökopista koostuva rakenne toistuu myös kiteessä ja täyttää
koko avaruuden. Tällaisen yksikkökopin sisällä on kuitenkin enemmän atomeita,
joten yhtälö (\ref{atompos}) antaa täsmälleen samat atomien paikat.

Sinkkivälkehila ja timanttihila ovat puolijohteissa yleisiä kidehiloja, jotka
ovat hyvin yksinkertaisia. Niiden yksikkökoppina on kuutio. Tämän kuution sivua
eli hilavektoreiden pituutta kutsutaan hilavakioksi. Sinkkivälkehila koostuu
kahdesta erilaisesta atomista. Timanttihila on muuten samanlainen, mutta siinä
nämä kaksi atomia ovat samoja. Kolmas puolijohteissa yleinen kidehila on
wurtsiittihila, jonka rakenne näyttää tietystä suunnasta katsottuna
kuusikulmaiselta. Tämänkin hilan yksikkökoppi voidaan kuitenkin esittää
kuusikulmaiselta näyttävästä rakenteestaan huolimatta suuntaissärmiönä. 

\begin{figure}[h!]
\igraph{0.6}{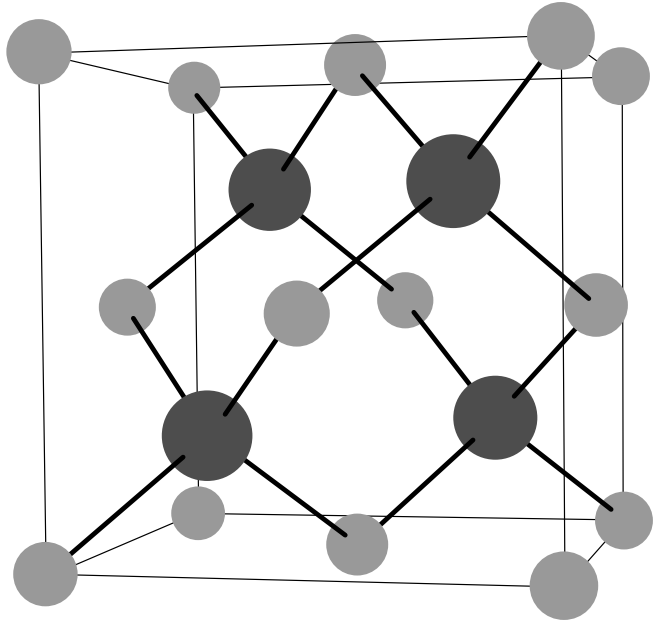}
\caption{Sinkkivälkehilan rakenne. Isoilla tummanharmailla ja pienillä
vaaleanharmailla palloilla kuvataan atomeita. Atomien sidokset on merkitty
paksuilla viivoilla. Sidoksia atomeihin, jotka eivät näy kuvassa, ei ole
merkitty. Yksi atomi on aina sitoutunut neljään erityyppiseen atomiin. Kuvassa
on yhden yksikkökopin sisällä tai reunalla olevat atomit. Yksikkökoppi on
ohuilla viivoilla rajattu alue.}
\label{zincblende}
\end{figure}

Kuvassa \ref{zincblende} on esitetty sinkkivälkehilan rakenne. Kuvassa atomeita on
merkitty harmailla ja mustilla palloilla. Kuvassa harmaat atomit ovat
yksikkökopin kulmissa ja yksikkökopin tahkojen keskellä. Mustat atomit ovat
sijoittuneet samalla tavalla, mutta mustien atomien hilaa on siirretty jokaisen
hilavektorin suuntaan neljänneshilavektorin verran. Yksikkökoppi olisi
tietenkin voitu valita myös niin, että mustat atomit olisivat yksikkökopin
tahkoilla ja kulmissa ja harmaat atomit sisällä. Sekä mustia että harmaita
atomeita on yksikkökopissa neljä. Kuvaan on piirretty 14 harmaata atomia, mutta
niistä vain neljä kuuluu tähän yksikkökoppiin. Koska yksikkökopit toistuvat
kaikissa avaruuden ulottuvuuksissa, kahdeksan harmaata atomia on jokainen
kahdeksan eri yksikkökopin kulmassa. On valittava käytäntö, joka määrittää,
mihin näistä kahdeksasta yksikkökopista kulmassa oleva atomi kuuluu.
Vastaavasti tahkon keskellä olevat harmaat atomit kuuluvat vain toiseen niistä
yksikkökopeista, joiden reunalla ne ovat.

Koska atomit ovat sijoittuneet toistuvaan rakenteeseen, kiteen mikroskooppiset
ominaisuudet ovat jaksollisia paikkamuuttujan funktioita. Röntgendiffraktion
kannalta kiinnostavin ominaisuus on sähköinen
suskeptiivisuus $\chi(\vect{r})$, joka on skalaariarvoinen funktio. Sähköinen
suskeptiivisuus kuvaa aineen ja sähkökentän välistä vuorovaikutusta.
Röntgensäteilyn taajuuksilla magneettikentän ja aineen välinen vuorovaikutus on
niin heikko, ettei sitä tarvitse ottaa huomioon. Funktion $\chi(\vect{r})$ jaksoja
ovat hilavektorit $\vect{a_1}$, $\vect{a_2}$ ja $\vect{a_3}$
yhtälön (\ref{period}) mukaisesti.

\begin{equation}
\chi(\vect{r}) = \chi(\vect{r} + h\vect{a_1} + k\vect{a_2} + l\vect{a_3}), \condspace \forall h, k, l \in \mathbb{Z} \label{period}
\end{equation}

Yksiulotteinen jaksollinen funktio $f(x)$ voidaan esittää Fourier-sarjan avulla
kompleksisten eksponenttifunktioiden äärettömänä summana yhtälön
(\ref{fourierone}) mukaisesti. Yhtälössä $H$ on reaaliluku, joka käy läpi kaikki
arvot, jolla yksiulotteisen funktion jakso $a$ on myös kompleksisen
eksponenttifunktion jakso. Tämä ehto voidaan esittää yhtälönä
(\ref{conditionone}). Fourier-kertoimet $f_H$ lasketaan yhtälöstä
(\ref{coeffone}), jossa integraali otetaan yhden jakson yli ja jossa $l$ on
jakson pituus.

\begin{eqnarray}
f(x) & = & \sum_{H} f_H \exp(\I H x) \label{fourierone} \\
\exp(\I H h a) & = & 1, \condspace \forall h \in \mathbb{Z} \label{conditionone} \\
f_H & = & \frac{1}{l} \int f(x) \exp(-\I H x) \D{x} \label{coeffone}
\end{eqnarray}

Fourier-sarja on yleistettävissä kolmeen ulottuvuuteen, jolloin funktio
$\chi(\vect{r})$ voidaan myös esittää Fourier-sarjana. Kolmiulotteisessa
Fourier-sarjassa skalaari $H$ korvataan vektorilla $\vect{h}$, ja kompleksisessa eksponenttifunktiossa
reaalilukujen tulo korvataan skalaaritulolla. Funktio $\chi(\vect{r})$ saadaan
summasta (\ref{fourierthree}), jossa $\vect{h}$ käy läpi yhtälön
(\ref{conditionthree}) mukaiset arvot. Sarjan kertoimet $\chi_\vect{h}$ saadaan
yhtälöstä (\ref{coeffthree}), jossa tilavuusintegraali lasketaan yksikkökopin
yli ja jossa $V$ on tämän yksikkökopin tilavuus.

\begin{eqnarray}
\chi(\vect{r}) & = & \sum_{\vect{h}} \chi_{\vect{h}} \exp(\I {\vect{h}} \cdot {\vect{r}}) \label{fourierthree} \\
\exp(\I \vect{h} \cdot (h \vect{a_1} + k \vect{a_2} + l \vect{a_3})) & = & 1, \condspace \forall h, k, l \in \mathbb{Z} \label{conditionthree} \\
\chi_\vect{h} & = & \frac{1}{V} \iiint \chi(\vect{r}) \exp(-\I \vect{h} \cdot \vect{r}) \D{V} \label{coeffthree}
\end{eqnarray}

Ongelmaksi jää niiden vektorin $\vect{h}$ arvojen selvittäminen, joilla ehto
(\ref{conditionthree}) pätee. Tätä vektorien $\vect{h}$ joukkoa kutsutaan
käänteishilaksi. Kompleksinen eksponenttifunktio saa arvon 1 vain, kun sen
argumentti on $2\pi$:n moninkerta. Skalaaritulo ja hilavektorien kokonaislukukertoiminen
lineaarikombinaatio voidaan kirjoittaa matriisimuotoon
\begin{equation}
\trans{\vect{h}}\left[\vect{a_1} \vect{a_2} \vect{a_3}\right]\trans{\left[h k l\right]} = 2 \pi m,\condspace m \in \mathbb{Z},\ \forall h, k, l \in \mathbb{Z} \label{invlattprooffirst}
\end{equation}
joka toteutuu jos ja vain jos käänteishilan vektorin transpoosin ja
hilavektorimatriisin matriisitulon rivit ovat $2\pi$:n moninkertoja.
Ehto saadaan siis muotoon
\begin{equation}
\trans{\vect{h}}\left[\vect{a_1} \vect{a_2} \vect{a_3}\right] = 2 \pi \left[n_1 n_2 n_3\right],\condspace n_1, n_2, n_3 \in \mathbb{Z}
\end{equation}
josta voidaan ratkaista $\vect{h}$, jolle saadaan yhtälö
\begin{equation}
\vect{h} = 2\pi\trans{\left(\left[\vect{a_1} \vect{a_2} \vect{a_3}\right]^{-1}\right)} \trans{\left[n_1 n_2 n_3\right]}, \label{invlattprooflast}
\end{equation}
mikä osoittaa, että käänteishila on matemaattiselta olemukseltaan samanlainen
kuin hila, eli käänteishilan vektorit ovat kolmen vektorin
kokonaislukukertoimisia lineaarikombinaatioita. Näitä vektoreita kutsutaan
käänteishilavektoreiksi $\vect{b_1}$, $\vect{b_2}$ ja $\vect{b_3}$, ja ne
saadaan yhtälöstä
\begin{equation}
\left[\vect{b_1} \vect{b_2} \vect{b_3}\right] = 2\pi\trans{\left(\left[\vect{a_1} \vect{a_2} \vect{a_3}\right]^{-1}\right)}. \label{invlattvect}
\end{equation}

Yhtälöstä (\ref{invlattvect}) saadaan hilavektorien ja käänteishilavektorien
pistetulokaavat (\ref{pistetulo1}--\ref{pistetulo2}).

\begin{eqnarray}
\vect{a_i} \cdot \vect{b_i} & = & 2\pi \label{pistetulo1} \\
\vect{a_i} \cdot \vect{b_j} & = & 0, \ i \neq j \label{pistetulo2}
\end{eqnarray}

Käänteishila on käyttökelpoinen työkalu kaikkialla, missä tarvitaan sellainen
vektori, jonka pistetulo paikkavektorin kanssa antaa yksiköttömän luvun.
Fourier-sarjan lisäksi tällaista vektoria tarvitaan tasojen esittämiseen.
Avaruuden taso on yksinkertaisinta esittää normaalivektorinsa avulla, jolloin
tason yhtälö on muotoa $\vect{h} \cdot \vect{r} = C$,~$C$~vakio. Kide halkeaa
yleensä sellaisten tasojen suuntaisesti, joissa atomit ovat sijoittuneet
mahdollisimman yksinkertaisesti. Nämä tasot leikkaavat yksikkökopin kohdissa
$\vect{a_1}/h$, $\vect{a_2}/k$ ja $\vect{a_3}/l$, joissa $h$, $k$ ja $l$ ovat
pieniä kokonaislukuja. Hilavektorien ja käänteishilavektorien
pistetulokaavoista (\ref{pistetulo1}--\ref{pistetulo2}) nähdään, että tällaisen
tason normaalivektori on $\vect{h} = h\vect{b_1} + k\vect{b_2} + l\vect{b_3}$, kun
valitaan $C = 2\pi$. Jos näitä samansuuntaisia tasoja kuvitellaan olevan
avaruudessa tasavälein, siten että yksi taso on origossa ($C = 0$) ja toinen
on tämä äsken kuvattu taso ($C = 2\pi$), näiden tasojen väliselle etäisyydelle
saadaan yhtälö
\begin{eqnarray}
d & = & \frac{2\pi}{|\vect{h}|},
\end{eqnarray}
josta saadaan käänteishilavektorin pituudelle laskentakaava
\begin{eqnarray}
|\vect{h}| & = & \frac{2\pi}{d}. \label{reciprocallen}
\end{eqnarray}

Tasojen suunnat ilmoitetaan yleensä Millerin indeksien avulla, joita merkitään
$(hkl)$. Tämä tarkoittaa tasoa, jonka normaalivektori on $h\vect{b_1} +
k\vect{b_2} + l\vect{b_3}$. Jos indeksin yläpuolella on viiva, se tarkoittaa
että kyseinen indeksi on negatiivinen. Tason normaalivektori voidaan kertoa
mielivaltaisella vakiolla, joten esimerkiksi $(111)$ ja $(222)$ tarkoittavat
täsmälleen samaa tasoa. Samaa merkintää käytetään tosin myös muissa
yhteyksissä, joissa tarvitaan käänteishilan vektoreita. Esimerkiksi
röntgendiffraktion eri diffraktiokulmia vastaavat tasot voidaan esittää kätevästi Millerin
indekseillä. Tällöin kuitenkaan vektoreihin $(111)$ ja $(222)$ liittyvää diffraktiota
ei havaita samalla kulmalla, koska röntgendiffraktiossa käänteishilavektorin suunnan lisäksi
pituudella on merkitystä.

\subsection{Epitaktiset kerrosrakenteet \label{epitaxy}}

Epitaksia on ohuiden puolijohdekerrosten kasvatusmenetelmä, jossa alustakiteen
eli substraatin päälle kasvatetun kerroksen kiderakenne on jatkoa substraatin
kiderakenteelle. Epitaksia voidaan jakaa homo- ja heteroepitaksiaan.
Homoepitaksiassa kasvatetaan substraatin päälle samaa puolijohdetta kuin mistä
substraatti koostuu. Heteroepitaksialla tarkoitetaan epitaksiaa, jossa
substraatti ja kerros ovat eri puolijohteita. Epitaktisia kerroksia voidaan
kasvattaa myös useita päällekkäin.

\begin{figure}[t]
\igraph{0.6}{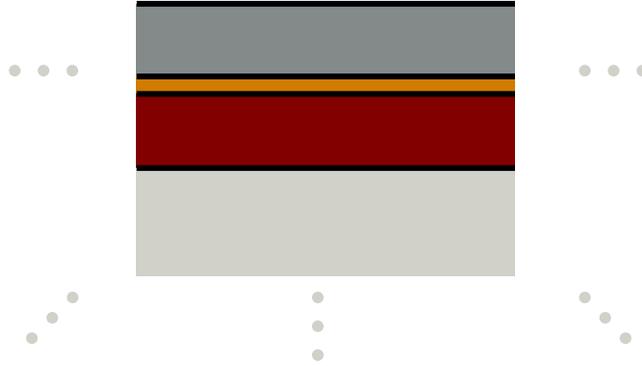}
\caption{Monikerrosrakenteen malli. Pohjimmaisin kerros eli substraatti oletetaan äärettömän paksuksi, ja kaikki kerrokset oletetaan äärettömän leveiksi ja syviksi. Ylimmäisen kerroksen yläpuolella on ilmaa.}
\label{multilayer}
\end{figure}

Tässä työssä tarkastellaan kuvan \ref{multilayer} kaltaisten rakenteiden
karakterisointia röntgenmittauksilla. Paksun substraatin päälle on kasvatettu
$N$ homogeenista kerrosta, joilla jokaisella on oma koostumus. Rakenne
oletetaan äärettömän leveäksi, ja substraatti oletetaan äärettömän paksuksi.
Kerrosten väliset rajapinnat oletetaan yhdensuuntaisiksi.
Röntgendiffraktiomittauksen tulosten analysoinnissa kerrosten oletetaan olevan
aina epitaktisia.

Seuraavaksi käsitellään substraatin päälle kasvatetun yhden heteroepitaktisen
kerroksen elastiikkaa. Käsittely on yleistettävissä monikerroksiseen
rakenteeseen, mutta tällöin käsittelyssä käytettävät substraatin $x$- ja
$y$-suuntaiset hilavakiot on korvattava kerroksen alapuolella olevan toisen
kerroksen hilavakioilla.

\begin{figure}[b]
\igraph{0.6}{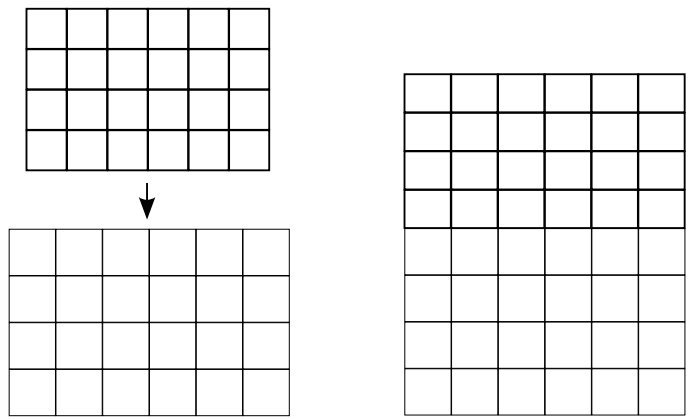}
\caption{Epitaktinen kerros venyy pinnan suuntaisesti, mihin liittyy vastakkaissuuntainen venymä kohtisuoraan pintaa vasten.}
\label{epikuva}
\end{figure}

Heteroepitaksiassa ohutkerroksen ja substraatin hilavakiot ovat erisuuruisia.
Koska heteroepitaktisen ohutkerroksen kiderakenne on jatkoa substraatin
kiderakenteelle, substraatin ja ohutkerroksen liitos saa ohutkerroksen
jännittyneeseen tilaan. Ohutkerros venyy niin, että sen pinnan suuntainen
hilavakio on sama kuin substraatilla. Substraatti ei veny, koska se on paljon
ohutkerrosta paksumpi. Kuvassa \ref{epikuva} on havainnollistettu kerroksen
venymistä. Tässä kuvassa ohutkerros venyy vaakasuunnassa (ja syvyyssuunnassa).

Venymistä pinnan suunnassa vastaa vastakkaissuuntainen venyminen kohtisuoraan.
Tämä kohtisuora venyminen saadaan laskettua ohutkerroksen elastisista
vakioista. Pystysuunnassa tiedetään, että jännitystä $\sigma_{zz}$ ei ole.
Pystysuuntainen venymä $\epsilon_{zz}$ halutaan selville. Pinnan suuntaiset
venymät $\epsilon_{xx}$ ja $\epsilon_{yy}$ saadaan laskettua ohutkerroksen
todellisesta ja jännittymättömästä hilavakiosta. Leikkausjännitystä tai
leikkausvenymää ei ole.

Venymät $x$-suunnassa määritellään
\begin{eqnarray}
\epsilon_{xx} & = & \frac{d_x - d_{x,0}}{d_{x,0}}, \label{strain}
\end{eqnarray}
missä $d_x$ on todellinen hilavakio $x$-suunnassa ja $d_{x,0}$ on hilavakio tässä
suunnassa venymättömässä tilassa. Vastaava yhtälö pätee tietysti muihinkin
suuntiin. Ideaalisella kerroksella $d_x$ ja $d_y$ ovat ohutkerrokselle samat kuin
substraatille. \cite[s.~18]{XSS}

Epäideaalinen kerros ei ole välttämättä täysin venynyt $x$- ja $y$-suunnissa niin, että
sen hilavakio olisi sama kuin substraatilla. Tällaisen kerroksen epäideaalisuutta
voidaan kuvata relaksaatiolla, joka määritellään yhtälöllä
\begin{eqnarray}
\epsilon_{xx} & = & \epsilon_{xx,ideal} (1 - R_x),
\end{eqnarray}
jossa $\epsilon_{xx,ideal}$ on ideaalisen kerroksen venymä, $\epsilon_{xx}$ on
kerroksen oikea venymä ja $R_x$ on relaksaatio $x$-suunnassa. Myös $y$-suunnan relaksaatio
määritellään vastaavasti. \cite[s.~18]{XSS}

Venymä $z$-suunnassa saadaan laskettua $x$- ja $y$-suuntaisten venymien avulla
lineaarisen elastiikan laeista, joista voidaan johtaa yhtälö
\begin{eqnarray}
\epsilon_{zz} & = & \frac{-\nu}{1-\nu} (\epsilon_{xx} + \epsilon_{yy}),
\end{eqnarray}
jossa $\nu$ on aineen Poissonin vakio kasvatussuuntaan. Vakio voi riippua suunnasta.
\cite[s.~16]{XSS}

Kun venymä $z$-suunnassa tiedetään, yhtälöstä (\ref{strain}) voidaan ratkaista
$z$-suuntainen hilavakio, jolle saadaan yhtälö
\begin{eqnarray}
d_z & = & d_{z,0}(1 + \epsilon_{zz}).
\end{eqnarray}

Useasta yhdisteestä koostuvan kerroksen hilavakio voidaan laskea Vegardin
laista. Tämän lineaarisen lain mukaan moniyhdistekerroksen hilavakio on yhdisteiden
suhteellisilla pitoisuuksilla painotettu keskiarvo niiden hilavakioista. Poikkeama lineaarisuudesta
on yleensä pieni ja hankala mitata. Jos yhdisteiden
suhteelliset pitoisuudet ovat $p_i$ ja hilavakiot $d_i$, moniyhdistekerroksen hilavakio
on siis \cite[s.~204]{XSS}
\begin{eqnarray}
d & = & \sum_{i} p_i d_i.\label{vegard}
\end{eqnarray}
Myös Poissonin luvulle käytetään vastaavaa lineaarista lakia
\begin{eqnarray}
\nu & = & \sum_{i} p_i \nu_i.\label{vegardpoisson}
\end{eqnarray}

\subsection{Aineen vuorovaikutus röntgensäteilyn kanssa \label{susc}}

Aineen suskeptiivisuuden laskeminen röntgentaajuuksilla perustuu
yksinkertaiseen atomimalliin, jossa atomin positiivisesti varautuneen ytimen
ympärillä on elektronipilveä kuvaava yhtäsuuri negatiivinen varausjakauma. Jos
atomi ei ole sähkökentässä, varausjakauman keskiarvo on atomiytimen kohdalla,
joten atomilla ei ole dipolimomenttia. Kun atomi asetetaan sähkökenttään,
elektronipilvi siirtyy siten, että varausjakauman keskiarvo ei enää ole
atomiytimen kohdalla. Syntyy ulkoista kenttää vastustava dipolimomentti.

Atomia mallinnetaan lineaarisella toisen kertaluvun differentiaaliyhtälöllä.
Yhtälö perustuu klassiseen fysiikkaan ja on siten tarkkaan ottaen virheellinen,
mutta auttaa ymmärtämään, mistä suskeptiivisuus johtuu\cite[s.~165]{cryst}.
Varausjakaumaan kohdistuva nettovoima oletetaan lineaariseksi poikkeaman
funktioksi, mikä pätee pienillä poikkeamilla. Mukaan otetaan myös
varausjakauman siirtymisnopeuteen verrannollinen vaimennustermi. Jos
elektronipilvessä on vain yksi elektroni, liikeyhtälö on
\begin{eqnarray}
m_e \Df{^{2}\vect{x}}{t^2} & = & -2m_e\gamma \Df{\vect{x}}{t} - m_e \omega_0^2\vect{x} - e\vect{E}, \label{liikeyht}
\end{eqnarray}
missä $m_e$ on elektronin massa, $e$ on alkeisvaraus, $2m_e\gamma$ on vaimennustermi,
josta kaksi kertaa elektronin massa on otettu omaksi tekijäkseen,
$m_e \omega_0^2$ on elektronipilveen kohdistuvan nettovoiman ja poikkeaman
verrannollisuuskerroin, josta elektronin massa on otettu tekijäksi ja
loput esitetään neliönä, $\vect{E}$ on sähkökenttä ja $\vect{x}$ on varausjakauman poikkeama
normaalitilasta.

\begin{figure}[t]
\igraph{0.5}{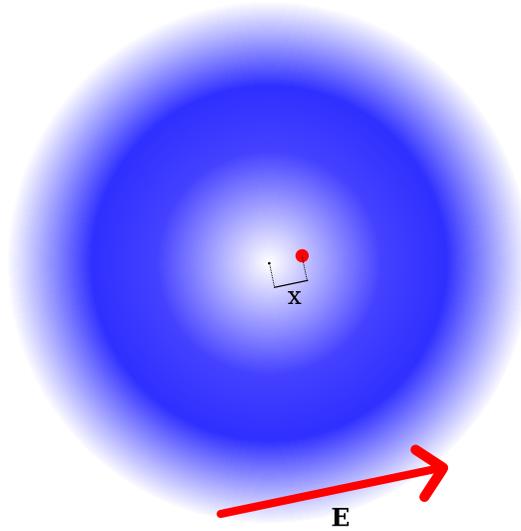}
\caption{Atomimalli. Positiivisesti varautuneen ytimen (punainen pallo)
ympärillä on negatiivisesti varautunut elektronipilvi. Tavallisesti elektronipilven
varausjakauman keskipiste on ytimen kohdalla, mutta sähkökenttä $\vect{E}$ siirtää
varausjakauman keskipistettä etäisyydelle $\vect{x}$ atomiytimestä, mikä indusoi atomiin dipolin.}
\end{figure}

Koska sähkömagneettisessa säteilyssä kenttä värähtelee sinimuotoisesti,
yhtälöön (\ref{liikeyht}) sijoitetaan $\vect{E} = \vect{E_0} \exp(\I \omega t)$. Aikaharmonisia
kenttiä käyttämällä kentän ja poikkeaman vaihe-eroa on helpompi
käsitellä, koska tällöin $\vect{x}$ on verrannollinen $\vect{E}$:hen. Sinimuotoinen
kenttä voidaan tietysti esittää kahden aikaharmonisen kentän
summana, jossa yhden eksponenttifunktion taajuus on positiivinen ja toisen
negatiivinen. Ratkaisuksi saadaan\cite[s.~35]{XSS}
\begin{eqnarray}
\vect{x} & = & -\frac{e}{m_e(\omega_0^2 + 2\I\gamma\omega - \omega^2)} \vect{E} \label{ratkaisu}.
\end{eqnarray}

Röntgensäteillä taajuus $\omega$ on suuri, joten ratkaisun (\ref{ratkaisu}) nimittäjän voidaan
approksimoida olevan $-m_e\omega^2$. Approksimaatio aiheuttaa pienen virheen
reaaliosaan ja unohtaa kokonaan imaginääriosan. Approksimaation tekemällä elektronin
polarisaatioksi $\vect{P} = -\vect{x}e\rho_e(\vect{r})$ saadaan
\begin{equation}
\vect{P} = -\frac{e^2}{m_e\omega^2}\rho_e(\vect{r})\vect{E} = -\frac{e^2{\lambda}^2}{4\pi^{2}m_ec^2}\rho_e(\vect{r})\vect{E}, \label{polarization}
\end{equation}
jossa $\lambda$ on röntgensäteilyn aallonpituus, $c$ on valon nopeus ja $\rho_e(\vect{r})$ on elektronitiheys. Yhtälö (\ref{polarization}) pätee tietysti myös protoneille, mutta koska protonin massa on noin 1800-kertainen elektronin massaan nähden, riittää että pelkästään elektronit otetaan huomioon kokonaispolarisaatiota laskettaessa\cite[s.~37]{XSS}.

Sähköinen suskeptiivisuus $\chi$ on polarisaation ja sähkökentän yksikötön
verrannollisuuskerroin, jonka määritelmänä on yhtälö $\vect{P} = \epsilon_0
\chi \vect{E}$. Yhtälöstä (\ref{polarization}) saadaan suskeptiivisuudelle
yhtälö
\begin{equation}
\chi = -\frac{e^2{\lambda}^2}{4\pi^{2}\epsilon_{0}m_ec^2}\rho_e(\vect{r})  = -\frac{r_e{\lambda}^2}{\pi}\rho_e(\vect{r}), \label{susc1}
\end{equation}
jossa
\begin{equation}
r_e = \frac{1}{4\pi\epsilon_0}\frac{e^2}{m_ec^2} \approx 2{,}81794\cdot10^{-15} \us{m}
\end{equation}
on luonnonvakio, jota kutsutaan klassiseksi elektronin säteeksi. Sama yhtälö on lähteessä \cite[s.~37]{XSS}, mutta
siinä käytetään SI-järjestelmän sijasta CGS-järjestelmää.

Olettaen, että atomit ovat pistemäisiä, elektronitiheydelle saadaan esitys
\begin{eqnarray}
\rho_e(\vect{r}) & = & \sum_{\vect{r}} Z_{\vect{r}} \delta(\vect{r}), \label{rhoe}
\end{eqnarray}
jossa $Z_{\vect{r}}$ on paikassa $\vect{r}$ olevan atomin elektronien määrä
ja $\delta(\vect{r})$ on kolmiulotteinen Diracin deltafunktio. Yhtälöt (\ref{susc1}) ja (\ref{rhoe}) voidaan yhdistää ja sijoittaa
luvussa (\ref{crystals}) esitettyyn Fourier-integraaliin yksikkökopin yli (\ref{coeffthree}). Suskeptiivisuuden
Fourier-kertoimiksi saadaan
\begin{eqnarray}
\chi_{\vect{h}} & = & -\frac{r_e {\lambda}^2}{\pi V} \sum_{\vect{r}} Z_{\vect{r}} \exp(-\I\vect{h}\cdot\vect{r}), \label{suscZ}
\end{eqnarray}
jossa summa lasketaan kaikkien yksikkökoppiin kuuluvien atomien yli.

Todellisuudessa $\chi_{\vect{h}}$ on kuitenkin hieman monimutkaisempi. Yhtälön
(\ref{ratkaisu}) nimittäjää approksimoitiin, mikä jättää suskeptiivisuuden
imaginääriosan huomiotta ja muuttaa reaaliosaa hieman. Tämän approksimaation
tarkkuus riippuu röntgensäteiden taajuudesta. Tietyillä taajuuksilla
approksimaatio ei päde lainkaan, mitä kutsutaan anomaaliseksi dispersioksi\cite[s.~165]{cryst}.
Atomit eivät myöskään ole
pistemäisiä, vaan atomiytimen ympärillä on elektronipilvi, jolla on tietty
varausjakauma. Varausjakauma riippuu myös lämpötilasta.
Pistevarausapproksimaation tarkkuus riippuu käänteishilan vektorista
$\vect{h}$. Fourier-sarjassa suuret käänteishilan vektorit vastaavat pieniä
paikkaresoluutioita, joten suurilla käänteishilan vektoreilla
pistevarausapproksimaatio on epätarkka. Lisäksi on huomioitava, että malli jolla
yhtälö (\ref{ratkaisu}) johdettiin perustuu klassiseen fysiikkaan.

Nämä ilmiöt voidaan ottaa huomioon korvaamalla atomiluku $Z_{\vect{r}}$
sirontakertoimella $F_{\vect{r}}(\omega)$, joka kerrotaan lämpötilariippuvuuden
huomioivalla Debye-Waller-kertoimella. Sirontakertoimen elektronien
varausjakaumasta riippuvaa osuutta laskettaessa yhtälön (\ref{coeffthree})
Fourier-integraali kiteen kokonaiselektronitiheydestä yksikkökopin yli
muunnetaan yhden yksikkökopin sisältämien atomien elektronitiheyden
Fourier-muunnokseksi koko avaruuden yli. Yksittäisen atomin elektronitiheyden
Fourier-muunnosta $f(\vect{h})$ kutsutaan atomin muotokertoimeksi\cite[s.~147]{cryst}.
Sirontakertoimeen lisätään myös taajuus- eli aallonpituusriippuvaiset atomikohtaiset termit,
joita kutsutaan Hönlin korjaustermeiksi\cite[s.~165]{cryst}.

Lähteessä \cite{Waasmaier} atomien muotokertoimet eli Fourier-muunnokset $f(\vect{h})$ on parametrisoitu
approksimoimalla niitä funktioilla
\begin{eqnarray}
f(s) & = & \sum_{i=1}^{5} a_i \exp(-b_i s^2) + c, \label{formfactor}
\end{eqnarray}
jossa parametri $s = |\vect{h}| / (4\pi) = \sin(\theta_B)/\lambda$. Lähteessä \cite{Waasmaier} käytetään
jälkimmäistä muotoa parametrille $s$, mutta luvussa \ref{xrd} johdetun Braggin lain avulla
voidaan molemmat muodot osoittaa samoiksi. Atomikohtaiset parametrit $a_i$, $b_i$ ja $c$ määrittävät siis 
atomin elektronitiheyden Fourier-muunnoksen.

Lähteessä \cite{Henke} on taulukoitu sirontakertoimen reaali- ja imaginääriosien
$f_1(\lambda)$ ja $f_2(\lambda)$ vakiokomponentit. Koska Hönlin korjaustermien riippuvuus
diffraktiokulmasta eli käänteishilan vektorista on hyvin pieni\cite[s.~166]{cryst}, lähteen \cite{Henke} arvoja voidaan käyttää myös
röntgendiffraktiossa tarvittavien suskeptiivisuuden muiden Fourier-komponenttien laskennassa.
Koska vakiokertoimelle $s = 0$, voidaan
sirontakerroin ilmaista kahdella eri tavalla
\begin{eqnarray}
F(\lambda) = f(0) + f'(\lambda) + \I f''(\lambda) = f_1(\lambda) + \I f_2(\lambda),
\end{eqnarray}
joista saadaan kaavat lähteen \cite{Henke} taulukkoarvojen muuntamiseksi Hönlin
korjaustermeiksi
\begin{eqnarray}
f'(\lambda) & = & f_1(\lambda) - f(0) \label{hoenlconv1} \\
f''(\lambda) & = & f_2(\lambda) \label{hoenlconv2}.
\end{eqnarray}

Elektronien lämpöliike muuttaa varausjakaumaa, mikä otetaan huomioon kertomalla
sirontakerroin $F(\lambda)$ Debye-Waller-lämpötilakertoimella $\exp(-Bs^2)$,
jossa $s$ on määritelty samalla tavalla kuin muotokertoimen laskennassa. Debye-Waller-lämpötilakerroin
perustuu oletukseen, että varausjakauma leviää pallosymmetrisen gaussisen
jakauman mukaisesti\cite[s.~149]{cryst}. Tässä $B$ on Debye-Waller-$B$-kerroin,
jonka arvoja on taulukoitu lähteessä \cite{Peng}. Jos $B$-kertoimia ei ole
taulukossa, ne voidaan laskea Debyen lämpötilasta\cite{Peng}. Debyen
$B$-kertoimia ei tunneta kaikille alkuaineille, ja ne riippuvat atomin lisäksi
siitä kiteestä, jossa atomi on\cite{Peng}. Tämä johtuu siitä, että atomin
lämpöliike riippuu sen ympärillä olevista atomeista\cite[s.~148]{cryst}. Jos
alkuaineelle on taulukoitu useita $B$-kertoimia erilaisille kiteille, arvoista
käytetään keskiarvoa, mitä ehdotetaan lähteessä \cite{polarizabilities}.

Jos aine on seos, yhdellä paikalla voi olla yksi useammasta atomista.
Esimerkiksi Al$_{0{,}2}$Ga$_{0{,}8}$As on III- ja V-ryhmien atomeista koostuva
puolijohde, jonka kidehila on sinkkivälkehila. Tämän aineen kiderakenteessa
III-ryhmän atomin paikalla on todennäköisyydellä $p = 0{,}2$ alumiiniatomi
ja todennäköisyydellä $p = 0{,}8$ galliumatomi. Todennäköisyyttä $p$ kutsutaan
okkupaatioluvuksi\cite[s.~87]{cryst}. Suskeptiivisuusyhtälössä (\ref{suscZ})
otetaan summaan sekä gallium- että alumiiniatomit, mutta nämä termit kerrotaan
galliumin ja alumiinin okkupaatioluvuilla.

Lopullinen laskentakaava suskeptiivisuudelle on siis
\begin{eqnarray}
\chi_{\vect{h}} & = & -\frac{r_e {\lambda}^2}{\pi V} \sum_{\vect{r}} p_\vect{r} \left( f_\vect{r}(s) + f_\vect{r}'(\lambda) + f_\vect{r}''(\lambda) \right) \exp(-Bs^2) \exp(-\I\vect{h}\cdot\vect{r}), \label{suscfin}
\end{eqnarray}

jossa summa lasketaan kaikkien yksikkökopin atomeiden yli, $V$ on yksikkökopin
tilavuus, $\vect{r}$ on atomin paikka, $p_\vect{r}$ on okkupaatioluku,
$r_e$ on klassinen elektronin säde, $\lambda$ on
röntgensäteilyn aallonpituus ja $s = |\vect{h}|/(4\pi)$. Yksikkökopin tilavuudessa $V$
ja käänteishilavektorin pituudessa $|\vect{h}|$ on otettava huomioon yksikkökopin venyminen (luku \ref{epitaxy}). Käytännössä
käänteishilavektorin pituus on yksinkertaisinta laskea kaavalla (\ref{reciprocallen}). Pistetulo $\vect{h}\cdot\vect{r}$
lasketaan esittämällä atomin paikka $\vect{r}$ hilavektorien reaalikertoimisena lineaarikombinaationa,
esittämällä $\vect{h}$ käänteishilavektorien kokonaislukukertoimisena lineaarikombinaationa ja käyttämällä
pistetulokaavoja (\ref{pistetulo1}--\ref{pistetulo2}).
Lähteessä \cite{Henke}
taulukoitujen arvojen perusteella saadaan atomikohtaiset aallonpituusriippuvat
Hönlin korjaustermit $f_\vect{r}'(\lambda)$ ja $f_\vect{r}''(\lambda)$, jotka
lasketaan kaavoista (\ref{hoenlconv1}--\ref{hoenlconv2}). Muotokerroinfunktio
$f_\vect{r}(s)$ on määritelty yhtälössä (\ref{formfactor}), jonka atomikohtaiset
taulukkoarvot $a_i$, $b_i$ ja $c$ saadaan lähteestä \cite{Waasmaier}.
$B$ on Debye-Waller-$B$-kerroin, joka otetaan joko suoraan taulukosta
tai lasketaan taulukoidun Debyen lämpötilan avulla\cite{Peng}.

\subsection{Röntgendiffraktio \label{xrd}}

Röntgendiffraktio (XRD) on ilmiö, jossa kiderakenteessa olevat säännöllisesti
järjestäytyneet atomit sirottavat säteilyä joka puolelle palloaaltona, joka
interferoi tietyissä suunnissa vahvistavasti. Näitä suuntia kutsutaan
diffraktiosuunniksi. Diffraktioilmiötä voidaan kuvata kahdella
diffraktioteorialla, jotka ovat kinemaattinen diffraktioteoria ja dynaaminen
diffraktioteoria. Kinemaattisessa diffraktioteoriassa oletetaan, että kiteeseen
saapuu tasoaalto, jonka intensiteetti pysyy vakiona. Näinhän ei tietenkään
oikeasti ole, sillä atomien sirottaman säteilyn interferenssi vaikuttaa myös
kiteeseen saapuvaan tasoaaltoon. Dynaamisessa diffraktioteoriassa
sirontailmiötä käsitellään sähköisen suskeptiivisuuden avulla, ja
diffraktioilmiö saadaan Maxwellin yhtälöiden ratkaisuista. Kinemaattinen
diffraktioteoria on tärkeä ymmärtää, koska myös dynaamisessa
diffraktioteoriassa tarvitaan kinemaattiseen diffraktioteoriaan liittyviä
käsitteitä. 

Oletetaan kinemaattisen diffraktioteorian mukaisesti, että kiteeseen saapuu aikaharmoninen tasoaalto
\begin{equation}
\vect{E_0} = \vect{A_0}\exp(-\I\vect{k_0}\cdot\vect{r})\exp(\I\omega t)
           = \vect{E_{0,0}}\exp(\I\omega t),
\end{equation}
jonka aikariippuvuus piilotetaan käsittelemällä aikaharmonisen kentän kerrointa.
$\vect{k_0}$ on säteilyn aaltovektori ja $\vect{A_0}$ on sähkökentän
värähtelyamplitudi. Kiteestä diffraktoituu tiettyihin suuntiin aikaharmonisia tasoaaltoja, joiden kertoimet ovat
\begin{eqnarray}
\vect{E_{1,0}} & = & \vect{A_1}\exp(-\I\vect{k_1}\cdot\vect{r}),
\end{eqnarray}
jossa aaltovektoreiden pituudet ovat samat, eli $|\vect{k_1}| = |\vect{k_0}|$, koska säteilyn taajuus ei muutu.

Oletetaan, että jokainen hilapiste heijastaa säteilyä tietyllä
heijastuskertoimella $k$ siten, että $|\vect{E_{j,1}}| = k|\vect{E_0}|$, jossa
$\vect{E_{j,1}}$ on kyseisestä hilapisteestä $r_j$ sironneen säteilyn osuus
lähtevästä tasoaallosta. Diffraktio tapahtuu, kun kaikista hilapisteistä
sironnut säteily on samassa vaiheessa tarkasteltuna esimerkiksi pisteessä
$\vect{r} = 0$.  Kun otetaan huomioon tulevan tasoaallon vaihesiirto origosta
pisteeseen $\vect{r_j}$, heijastuskerroin ja lähtevän säteilyn vaihesiirto
takaisin origoon, saadaan aikaharmonisen sirontakentän kertoimeksi origossa
\begin{equation}
\vect{E_{j,1,0}} = \vect{\hat{u}}|\vect{A_0}|\exp(-\I\vect{k_0}\cdot\vect{r_j}) k \exp(\I\vect{k_1}\cdot\vect{r_j})
             = \vect{\hat{u}}|\vect{A_0}|k\exp(\I(\vect{k_1}-\vect{k_0})\cdot\vect{r_j}),
\end{equation}
jossa $\vect{\hat{u}}$ on yksikkövektori, joka määrittää sähkökentän suunnan.

\begin{figure}[tb]
\igraph{0.7}{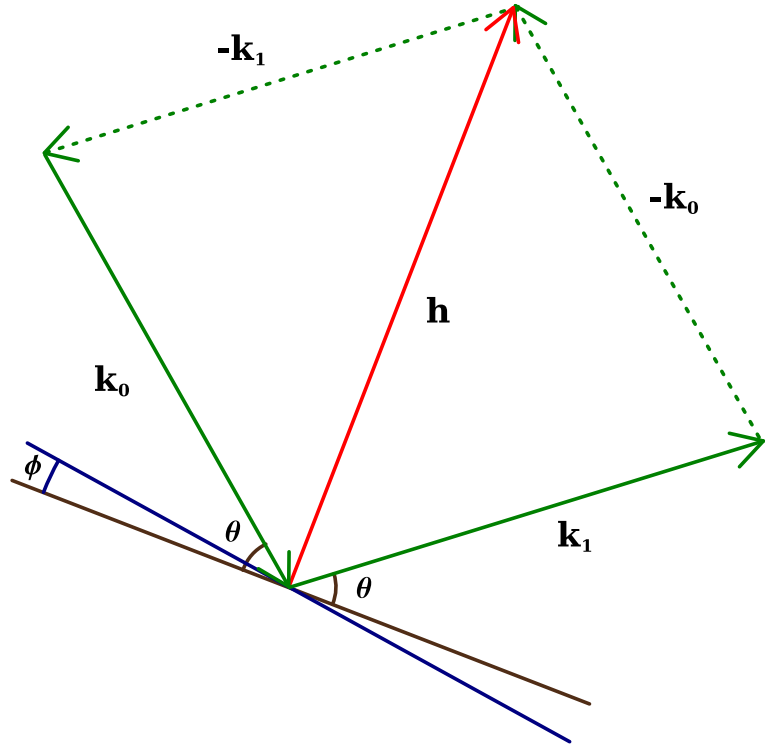}
\caption{Röntgendiffraktioon liittyvät tasot, kulmat ja vektorit. Vihreät vektorit $\vect{k_0}$ ja $\vect{k_1}$
ovat tulevan ja lähtevän säteen aaltovektorit. Myös vektorien vastavektorit ovat kuvassa.
Vektori $\vect{h} = \vect{k_1}-\vect{k_0}$ on käänteishilan vektori. Ruskea taso on diffraktiotaso,
eli taso jonka normaalivektori $\vect{h}$ on. Säteiden tulo- ja lähtökulmat suhteessa diffraktiotasoon ovat
samat, ja niitä merkitään $\theta$:lla. Näytteen pinta on sininen taso. Näytteen pinnan ja diffraktiotason
välinen kulma on $\phi$.}
\label{xrdkuva}
\end{figure}

Jotta säteet olisivat samassa vaiheessa, eksponenttitekijän täytyy olla
hilapisteestä $\vect{r_j}$ riippumaton vakio. Koska myös paikassa $2\vect{r_j}$
on hilapiste, vakion täytyy lisäksi olla 1. Ehto diffraktiolle on siis
\begin{eqnarray}
\exp(\I(\vect{k_1}-\vect{k_0})\cdot\vect{r_j}) & = & 1 \condspace \forall{\vect{r_j}},
\end{eqnarray}
mikä on suoraan käänteishilan määritelmä --- ehto toteutuu kaikilla
hilapisteillä $\vect{r_j}$ jos ja vain jos $\vect{k_1}-\vect{k_0}$ on jokin
käänteishilan vektori $\vect{h}$. Kuten aiemmin todettiin, lisäehtona vektorien
$\vect{k_1}$ ja $\vect{k_0}$ pituuksien on oltava yhtäsuuret.  Kuvassa
\ref{xrdkuva} havainnollistetaan diffraktiota. Tulevan ja lähtevän
aaltovektorin kulmat suhteessa diffraktiotasoon ovat samat. Tälle kulmalle
käytetään merkintää $\theta$. Diffraktioilmiö havaitaan, kun $\theta =
\theta_B$, jossa $\theta_B$ on Braggin kulma.

Kuvasta \ref{xrdkuva} saadaan diffraktiokulma $\theta_B$ trigonometrian avulla. Yhtälö
diffraktiokulman sinille on
\begin{eqnarray}
\sin \theta_B & = & \frac{|\vect{h}|}{2|\vect{k_0}|},
\end{eqnarray}
johon voidaan sijoittaa käänteishilan vektorin pituus $|\vect{h}| = \frac{2\pi}{d}$
ja aaltovektorin pituus $|\vect{k_0}| = \frac{2\pi}{\lambda}$. Yhtälö muuttuu lopulta
muotoon
\begin{eqnarray}
2d\sin\theta_B & = & \lambda, \label{bragg}
\end{eqnarray}
missä $\lambda$ on aallonpituus, $d$ on heijastavien tasojen välinen etäisyys
ja $\theta_B$ on Braggin kulma. Yhtälöä
(\ref{bragg}) kutsutaan Braggin laiksi. Braggin laissa heijastavat tasot ovat
täysin käänteishilan vektorin avulla määriteltyjä, joten Braggin laki pätee, vaikka
tasojen etäisyys jaettaisiin millä tahansa kokonaisluvulla $n$, sillä myös $n\vect{h}$ on
käänteishilan vektori.

Dynaaminen diffraktioteoria perustuu Maxwellin yhtälöihin, jotka ovat väliaineessa
\begin{eqnarray}
\nabla \times \vect{E} & = & -\Df{\vect{B}}{t} \\
\nabla \times \vect{H} & = & \vect{J}_{\mathrm{free}}+\Df{\vect{D}}{t} \\
\nabla \cdot \vect{B} & = & 0 \\
\nabla \cdot \vect{D} & = & \rho_{\mathrm{free}},
\end{eqnarray}
ja materiaalista riippuviin konstitutiivisiin yhtälöihin
\begin{eqnarray}
\vect{D} & = & \epsilon \vect{E} = \epsilon_0(1+\chi) \vect{E} \\
\vect{B} & = & \mu \vect{H} = \mu_0(1+\chi_m) \vect{H},
\end{eqnarray}
joissa $\vect{E}$ on sähkökentän voimakkuus ja $\vect{B}$ on magneettivuon tiheys.
Nämä ovat fysikaalisia sähkö- ja magneettikenttiä, jotka aiheuttavat sähkövaraukseen Lorentzin voiman.
Väliaineessa yhtälöiden ratkaisemista helpottaa, jos käyttää näiden kenttien lisäksi kenttiä
$\vect{D}$ ja $\vect{H}$, jotka ovat sähkövuon tiheys ja magneettikentän voimakkuus.
$\vect{J}_{\mathrm{free}}$ on vapaa virrantiheys, jossa ei ole otettu
huomioon polarisaation muutoksesta ja magnetisoitumisesta johtuvia
virrantiheyksiä. $\rho_{\mathrm{free}}$ on vapaa varaustiheys,
jossa ei ole otettu huomioon polarisaatiovaraustiheyttä. $\epsilon_0$ ja $\mu_0$ ovat
tyhjiön permittiivisyys ja permeabiliteetti, jotka ovat luonnonvakioita. $\chi$
on sähköinen suskeptiivisuus, joka kuvaa väliaineen polarisoitumista sähkökentän vaikutuksesta.
$\chi_m$ on magneettinen suskeptiivisuus, joka kuvaa väliaineen magnetisoitumista
magneettikentän vaikutuksesta.

Röntgendiffraktiossa vapaata varaustiheyttä $\rho_{\mathrm{free}}$ ei ole,
koska kaikki varaustiheys johtuu polarisaatiosta. Röntgentaajuuksilla
magneettinen suskeptiivisuus $\chi_m = 0$, eli väliaineen ja kenttien
vuorovaikutus röntgensäteillä on puhtaasti sähköinen. Myöskään vapaata
virrantiheyttä $\vect{J}_{\mathrm{free}}$ ei ole, koska kaikki virrantiheys
johtuu polarisaation aikamuutoksesta. $\chi$ on kiteisellä aineella paikan
funktio $\chi(\vect{r})$, joka voidaan esittää Fourier-sarjan avulla (luvut
\ref{crystals} ja \ref{susc}).

Tässä työssä tarkastellaan diffraktiota epitaktisista monikerrosrakenteista,
joissa diffraktio mitataan käänteishilavektorista, joka on kohtisuorassa
näytteen pintaa vastaan. Esimerkiksi jos näytteen pinnan Millerin indeksi on
($100$), diffraktio voisi tapahtua käänteishilavektoreista ($100$), ($200$), ($300$),
($400$) ja yleisesti ($n00$), jossa $n$ on mikä tahansa kokonaisluku. Sinkkivälke-
ja timanttihiloilla kuitenkin yksikkökopin sisäisestä rakenteesta johtuen diffraktioiden
($100$), ($200$) ja ($300$) intensiteetti on nolla tai hyvin pieni, joten näillä
kiderakenteilla käytetään vektoria ($400$). Näytteiden epäideaalisuudesta
johtuen käänteishilavektori ei itse asiassa ole aivan tarkalleen kohtisuorassa
pintaa vasten, vaan se poikkeaa kohtisuoruudesta pienen kulman $\phi$ verran.
Näytteeseen kohdistetaan röntgensäteitä kulmassa $\theta-\phi$ pinnan kanssa.
Diffraktoitunut säteily mitataan kulmassa $\theta+\phi$. Tällöin tulo- ja
lähtökulmat käänteishilan vektorin suhteen ovat samat, mikä on diffraktion
edellytys.

\begin{figure}[tb]
\igraph{1}{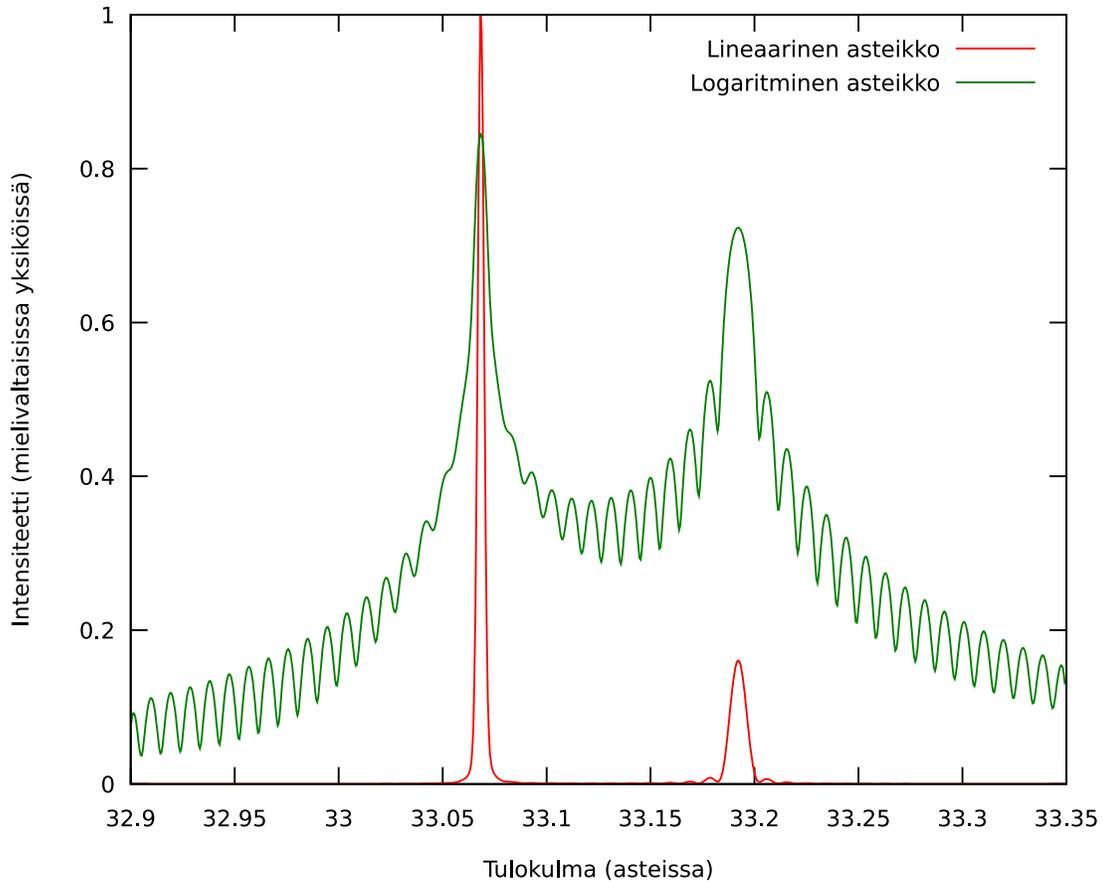}
\caption{Yksinkertaisen rakenteen simuloitu diffraktiointensiteetti
lineaarisella ja logaritmisella asteikolla. Lineaarisella asteikolla voimakkaat
diffraktiopiikit peittävät alleen aaltoilun, josta saadaan paksuusinformaatio.
Logaritmisella asteikolla aaltoilu erottuu hyvin.}
\label{scalecomp}
\end{figure}

Diffraktiointensiteettiä mitataan kulman $\theta$ funktiona. Jos näytteessä on
useita kerroksia, jokaisella kerroksella on oma Braggin kulmansa. Kun $\theta$
on tarkalleen yhtäsuuri kuin jonkin kerroksen Braggin kulma,
diffraktiointensiteetti on voimakas.  Näiden kulmien lähistöllä
diffraktiointensiteetti on huomattavasti pienempi, mutta sisältää paljon
tärkeää tietoa näytteen rakenteesta. Diffraktiointensiteetissä havaitaan
aaltoilua kulman funktiona, mistä saadaan selville kerrosten paksuudet.
Aaltoilu havaitaan parhaiten logaritmisella asteikolla, minkä vuoksi kuvaajia
on tapana käsitellä tällä asteikolla (kuva \ref{scalecomp}). Braggin
kulmista saadaan selville näytteiden koostumus, koska Braggin kulma riippuu
heijastavien tasojen välisestä etäisyydestä, joka on verrannollinen
hilavakioon. Hilavakio riippuu kerroksen koostumuksesta Vegardin lain
mukaisesti, ja myös kerroksen venyminen muuttaa hilavakioita eri suunnissa.

Yhtälöitä diffraktiointensiteetin laskentaan ei johdeta tässä, vaan yhtälöt ja
niiden ratkaisut esitetään tässä ainoastaan lyhyesti. Tarkempaa tietoa dynaamisesta
diffraktioteoriasta saa röntgendiffraktiokirjallisuudesta\cite{XSS}. Tässä
työssä lasketaan intensiteetti rekursiokaavoilla, jotka on johdettu
artikkelissa \cite{bartels}. Maxwellin yhtälöistä voidaan johtaa röntgendiffraktiolle
Takagi-Taupin-yhtälö olettaen, että vain sähköisen suskeptiivisuuden vakiokomponentilla $\chi_0$
ja Braggin kulmaa vastaavan käänteishilavektorin Fourier-komponenteilla $\chi_{\vect{h}}$ ja $\chi_{\vect{\overline{h}}}$ on merkitystä
diffraktiointensiteettiin. Tulevan ja lähtevän säteen amplitudien muutosta syvyyden
funktiona kuvaava Takagi-Taupin-yhtälö on
\pagebreak[0]
\begin{eqnarray}
-\I \Df{X}{T} & = & X^2 - 2\eta X + 1,
\end{eqnarray}
\pagebreak[0]
jossa
\pagebreak[0]
\begin{eqnarray}
X & = & \sqrt{\frac{\chi_{\vect{h}}}{\chi_{\overline{\vect{h}}}}} \sqrt{\left|\frac{\gamma_H}{\gamma_0}\right|} \frac{D_H}{D_0} \\
\eta & = & \frac{-b(\theta-\theta_B)\sin(2\theta_B) + \frac{1}{2}\chi_{0}(1-b)}{\sqrt{|b|}C\sqrt{\chi_{\vect{h}}\chi_{\overline{\vect{h}}}}} \\
T & = & \frac{\pi C t \sqrt{\chi_{\vect{h}}\chi_{\overline{\vect{h}}}} }{\lambda \sqrt{|\gamma_0\gamma_H|}} \\
b & = & \frac{\gamma_0}{\gamma_H} \\
\gamma_{0} & = & \sin(\theta_B-\phi) \\
\gamma_{H} & = & -\sin(\theta_B+\phi),
\end{eqnarray}

joissa $\theta$ on säteiden tulokulma suhteessa heijastaviin tasoihin,
$\theta_B$ on hilavektoria $\vect{h}$ vastaava Braggin kulma, $\phi$ on heijastavien
tasojen ja pinnan välinen kulma, $\chi_{\vect{h}}$
ja $\chi_{\vect{\overline{h}}}$ ovat näitä hilavektoreita vastaavat suskeptiivisuuden
Fourier-sarjan komponentit, $\chi_0$ on suskeptiivisuuden vakiokomponentti, $C$
on polarisaatiokerroin joka on $1$ $\sigma$-polarisaatiolle ja
$|\cos{2\theta_B}|$ $\pi$-polarisaatiolle, $\lambda$ on aallonpituus,
$\gamma_{0}$ ja $\gamma_{H}$ ovat Braggin kulmalla tulevan ja lähtevän säteen
suuntakulmien kosinit mitattuna pinnannormaalista, $b$ on heijastuksen asymmetrisyyttä kuvaava luku joka
on noin $-1$, $\eta$ on tulokulman ja Braggin kulman erotuksesta riippuva poikkeamaparametri,
$T$ on syvyyden $t$ avulla määritelty suure ja $X$ on tulevan ja
lähtevän säteiden amplitudien $D_H$ ja $D_0$ suhteen avulla määritelty suure. Suureet
$\eta$, $T$ ja $X$ ovat määritelty siten, että yhtälöstä tulisi mahdollisimman yksinkertainen.
Suskeptiivisuuden Fourier-komponentit lasketaan kaavalla (\ref{suscfin}), ja Braggin kulma saadaan yhtälöstä
(\ref{bragg}), jossa on otettava huomioon yksikkökopin venyminen (luku \ref{epitaxy}).

Yhtälö voidaan ratkaista separoimalla ja integroimalla. Jos $X$ on kerroksen
pohjassa $X_0$ ja kerroksen pinnalla $X_t$, ja kerros on homogeeninen ($\eta$
vakio), saadaan $X_t$ rekursiokaavan avulla
\begin{eqnarray}
X_t & = & \eta + \sqrt{\eta^2-1} \frac{S_1 + S_2}{S_1 - S_2} \label{ratk1},
\end{eqnarray}
jossa
\begin{eqnarray}
S_1 & = & (X_0 - \eta + \sqrt{\eta^2-1}) \exp(-\I T \sqrt{\eta^2-1}) \label{ratk2} \\
S_2 & = & (X_0 - \eta - \sqrt{\eta^2-1}) \exp(\I T \sqrt{\eta^2-1}). \label{ratk3}
\end{eqnarray}

Ratkaisukaavoissa (\ref{ratk1}-\ref{ratk3}) suure $T$ määritellään muuten saman
yhtälön avulla kuin aiemmin, mutta tässä yhtälössä $t$:llä on nyt erilainen
rooli. Aiemmin $t$ oli syvyysparametri ja $T$ tästä parametrista riippuva suure,
jonka suhteen derivoitiin. Nyt $t$ on kerroksen paksuus.

Pohjimmaisena kerroksena on huomattavasti muita kerroksia paksumpi substraatti.
Substraatti oletetaan tässä äärettömän paksuksi kerrokseksi, jolle voidaan rekursiokaavan
raja-arvosta johtaa Darwin-Prins-yhtälö
\begin{eqnarray}
X_{\infty} & = & \eta \pm \sqrt{\eta^2-1},
\end{eqnarray}
jossa merkki valitaan vastakkaiseksi kuin poikkeamaparametrin $\eta$ reaaliosan merkki.

Säteen heijastuvuus päällimmäisen kerroksen pinnalla saadaan suureen $X$
määritelmästä. Heijastuvuus ilmaistaan intensiteetille eikä kenttävoimakkuudelle.
Polarisoitumattoman säteen intensiteetti on $\sigma$- ja $\pi$-polarisoituneiden
säteiden intensiteettien keskiarvo.
Heijastuvuus ei ole suoraan kenttien amplitudien suhteen neliö, koska
siinä on otettava huomioon asymmetriaparametri $b$, joka muuttaa säteen
poikkileikkauksen suuruutta. Heijastuvuus on\cite{bartels}
\begin{eqnarray}
P & = & \left|\sqrt{\frac{\chi_{\vect{h}}}{\chi_{\vect{\overline{h}}}}}\right|\left|X\right|^2.
\end{eqnarray}

\clearpage
\section{Tutkimusongelma ja -menetelmät}

\subsection{Käänteisongelma \label{inverse}}

Jos näytteen rakenne tunnetaan, on helppo laskea röntgensäteilyn
diffraktiointensiteetti. Diffraktiointensiteetti lasketaan käyttäen
luvussa \ref{xrd} esitettyä dynaamista diffraktioteoriaa.

Näytteelle on rakennettava kerrosmalli, jonka kerrosten muuttujia ovat
röntgendiffraktion tapauksessa jokaisen kerroksen paksuus, koostumus ja
relaksaatio. Malliin kuuluu myös pinnan ja heijastavien tasojen välinen kulma.
Lisäksi tulevan röntgensäteilyn intensiteetti ja taustasäteilyn intensiteetti
otetaan malliin mukaan.

Merkitään mallin muuttujia $x_1, x_2, \ldots, x_{n}$. On siis olemassa funktio
$I$, jolle annetaan mallin muuttujat ja tulokulma ja joka laskee lähtevän
säteilyn intensiteetin näistä. Lisäksi tämän funktion arvo voidaan laskea
helposti. Merkitään tulokulmia $\alpha_k$:lla. Jokaista tulokulmaa kohti saadaan laskettua lähtevän säteilyn intensiteetti $I_{sim,k}$ yhtälöstä (\ref{laskuyht}).

\begin{equation}
I_{sim,k} = I(\alpha_k, x_1, x_2, \ldots, x_{n}) \label{laskuyht}
\end{equation}

Diffraktiomittauksissa kuitenkin intensiteetit ovat mitattuja suureita, ja
näiden tietojen avulla pitäisi saada laskettua muuttujien $x_i$ arvot. Jos
funktio on riittävän yksinkertainen, voitaisiin löytää menetelmä, jolla
riittävän monella tulokulmalla mitatuista intensiteeteistä voitaisiin laskea
muuttujien arvot. Intuitiivisesti $n$ tulokulman arvoa riittäisi, koska $n$
tuntematonta vaatii yleensä $n$ yhtälöä ratketakseen yksikäsitteisesti.

Yhtälön ratkaisemiseen perustuvassa menetelmässä on kuitenkin useita ongelmia.
Röntgendiffraktion mallinnukseen käytettävät yhtälöt ovat epälineaarisia, joten
niillä ei ole välttämättä yksikäsitteistä ratkaisua. Jos ratkaisu on, se olisi
etsittävä numeerisella iteraatiolla, mikä vaatisi alkuarvauksen. Useat
numeeriset ratkaisumenetelmät toimivat varmuudella vain, jos alkuarvaus on
riittävän lähellä ratkaisua. Jos ratkaisu löydettäisiin, siihen ei voitaisi
luottaa, koska jouduttaisiin olettamaan, että kerrosmalli on oikea.
Yhtälöryhmä myös saattaa olla sellainen, että pienikin kohina
mittausdatassa tai epäideaalisuus näytteessä tai mittausolosuhteissa vääristää
mittaustuloksia liikaa.

Käytännössä mittauksissa käytetään ylimääräytyvää yhtälöryhmää, jossa on
yhtälöitä huomattavasti enemmän kuin tuntemattomia. Yhtälöt ovat muotoa
(\ref{laskuyht}), mutta niissä tietysti käytetään mitattuja intensiteettejä
$I_{meas,k}$. Tällöin voidaan luottaa mitattuihin muuttujien arvoihin enemmän,
koska satunnaisesti valitulla virheellisellä mallilla laskettuja
intensiteettejä ei todennäköisesti saada vastaamaan mitattuja intensiteettejä.
Jos yhtälöitä on yhtä monta kuin tuntemattomia, mille tahansa mallille
löydetään ratkaisu riippumatta mallin totuudenmukaisuudesta.
Röntgendiffraktiokäyrät eivät kuitenkaan sisällä kaikkea mahdollista tietoa
näytteestä, joten virheellisen mallin voi saada sopimaan mittaustuloksiin.

Ylimääräytyvällä yhtälöryhmällä ei ole ratkaisua kuin äärimmäisen harvinaisissa
poikkeustapauksissa muun muassa mittausten kohinasta ja näytteiden
epäideaalisuudesta johtuen. Voidaan kuitenkin etsiä muuttujien $x_i$ arvot,
jolla lasketut intensiteetit vastaavat kokonaisuutena mahdollisimman hyvin
mitattuja intensiteettejä. Graafisesti tämä voidaan esittää lasketun
intensiteettikäyrän sovituksena mitattuun käyrään.  Käyrän sovitukseen
tarvitaan kelpoisuusfunktio, joka mittaa mitatun käyrän ja lasketun käyrän
eroa. Funktio mittaa itse asiassa käänteistä kelpoisuutta, sillä pienillä
kelpoisuusfunktioiden arvoilla mitatut intensiteetit eroavat vain vähän
lasketuista intensiteeteistä, eli kelpoisuus on hyvä.  Kelpoisuusfunktio laskee
siis mitattujen ja laskettujen intensiteettien perusteella näiden
intensiteettien eroavaisuuden, jota kuvataan reaaliluvulla.  Kelpoisuusfunktion
globaalista minimistä saadaan lopulliset mitatut arvot muuttujille $x_i$, eli
näytteen rakenteen selvittämiseksi on ratkaistava globaali optimointiongelma.

Kelpoisuusfunktiona tässä työssä käytetään logaritmisten intensiteettien
erotuksien neliösummaa (yhtälö \ref{fitness}). Erotuksien neliösumma on
perinteisesti käyrän sovituksessa käytetty kelpoisuusfunktio. Sovitus tehdään
logaritmisella asteikolla, koska tällaisella asteikolla diffraktioilmiön eri
yksityiskohdat tulevat parhaiten näkyvin (luku \ref{xrd}). Logaritmisen
asteikon käytöllä on myös huonoja ominaisuuksia, sillä logaritminen asteikko
voimistaa kohinaa asteikon alapäässä.

\begin{equation}
F = \sum_{k} \left(\log I_{meas,k} - \log I_{sim,k}\right)^2 \label{fitness}
\end{equation}

\subsection{Optimointialgoritmit \label{DE}}

Käänteisongelma ratkaistaan differentiaalievoluutiolla (DE, \emph{differential evolution}), joka on tehokas globaali optimointialgoritmi monimutkaisille funktioille.
Differentiaalievoluutiota on käytetty paljon useisiin käytännön
optimointiongelmiin, joissa sen on havaittu toimivan hyvin tehokkaasti.
Differentiaalievoluutiota on käytetty aiemmin myös 
röntgendiffraktiodatan analysoinnissa\cite{Wormington}. Sovitusalgoritmi on
kirjoitettu toimimaan Matlabilla ja GNU~Octavella.

Differentiaalievoluutio perustuu ainoastaan kelpoisuusfunktion arvon
laskemiseen toisin kuin monet muut algoritmit, kuten Newtonin iteraatio ja
gradienttilaskeutuminen, jotka tarvitsevat myös optimoitavan funktion
derivaattoja. Differentiaalievoluutio ei siis vaadi kelpoisuusfunktiolta
derivoituvuutta eikä edes jatkuvuutta. Kelpoisuusfunktion arvoa käytetään
ainoastaan vertailussa toisiin kelpoisuusfunktioiden arvoihin, joten jos
kelpoisuusfunktio on $F(x_1, \ldots, x_n)$ ja $f(x)$ on aidosti kasvava
funktio, niin myös $f(F(x_1, \ldots, x_n))$ toimii täsmälleen yhtä hyvin
kelpoisuusfunktiona. Differentiaalievoluutio on siis invariantti
kelpoisuusfunktion absoluuttiselle arvolle, ja ainoastaan suuruusjärjestyksellä
on merkitystä optimointiin. Tällainen invarianssi on hyödyllinen, sillä se
vähentää tarvetta kelpoisuusfunktion huolelliseen valintaan. Toisaalta se
vähentää mahdollisuutta optimoida kelpoisuusfunktio sellaiseksi, että
differentiaalievoluutio konvergoituu mahdollisimman hyvin tietyllä ongelmalla.
Voidaan siis todeta, että differentiaalievoluutio on hyvä yleiskäyttöisenä
optimointialgoritmina, jota kannattaa kokeilla tiettyyn globaaliin
optimointiongelmaan ensimmäisenä optimointialgoritmina. Muita vaihtoehtoisia
globaaleja optimointialgoritmeja ovat esimerkiksi Monte Carlo -menetelmä ja
simuloitu jäähdytys, mutta ne ovat tehottomia, koska ne eivät ota huomioon
kelpoisuusfunktion geometriaa\cite{Wormington}. Differentiaalievoluution
lisäksi on olemassa myös muita evoluutiopohjaisia optimointialgoritmeja.

Differentiaalievoluution perusidea on samankaltainen kuin biologisella
evoluutiolla. Differentiaalievoluutiossa optimointiparametrien $(x_1, \ldots,
x_n)$ arvoja kutsutaan yksilöksi. Yksittäistä optimointiparametria kutsutaan
geeniksi. Optimoinnin aikana ylläpidetään populaatiota, jossa yksilöitä on
vakiomäärä $N$. Tarkoituksena on maksimoida jokaisen populaation yksilön
kelpoisuus, eli minimoida kelpoisuusfunktion arvo. Populaation kelpoisuus
maksimoidaan mutaatiolla, risteytyksellä ja valinnalla. Mutaatiolla ja
risteytyksellä muodostetaan uusia yksilöitä, ja valinnalla rajoitetaan
kasvaneen populaation kokoa niin, että se pysyy vakiona $N$. Mutaation
ensisijaisena tarkoituksena on löytää uusia geenejä, jotka saattavat parantaa
kelpoisuutta. Suurin osa mutaatioista tietenkin heikentää kelpoisuutta, mutta
ilman mutaatiota uusia geenejä ei löydettäisi. Risteytyksen tarkoituksena on
muodostaa uusi yksilö, jolla on kahden vanhan yksilön hyvät geenit, mutta ei
huonoja. Suurin osa risteytyksistä tietysti ei paranna kelpoisuutta
merkittävästi, sillä risteytys saattaa ihan yhtä hyvin valita uuteen yksilöön
kahden vanhan yksilön huonot geenit. Geeneillä saattaa vieläpä olla
geenienvälisiä vuorovaikutuksia, jolloin esimerkiksi $(a_1, b_1)$ ja $(a_2,
b_2)$ saattavat olla hyviä yksilöitä, mutta $(a_1, b_2)$ ja $(a_2, b_1)$
huonoja. Ilman risteytystä ei kuitenkaan löydettäisi yhtä nopeasti yksilöitä,
jolla kaikki geenit ovat hyviä. Valinta lopulta poistaa uudesta populaatiosta
huonot yksilöt ja jättää hyvät jäljelle. Kun mutaatiota, risteytystä ja
valintaa kohdistaa populaatioon riittävän monta kertaa, populaatio lopulta
konvergoituu, ja löydetään jokin kelpoisuusfunktion minimi. Mitään takeita ei
ole siitä, että tämä minimi olisi globaali minimi, mutta käytännössä on
havaittu differentiaalievoluution löytävän globaalin minimin kohtuullisen
monimutkaisillakin funktioilla.

\begin{figure}[tb]
\igraph{0.7}{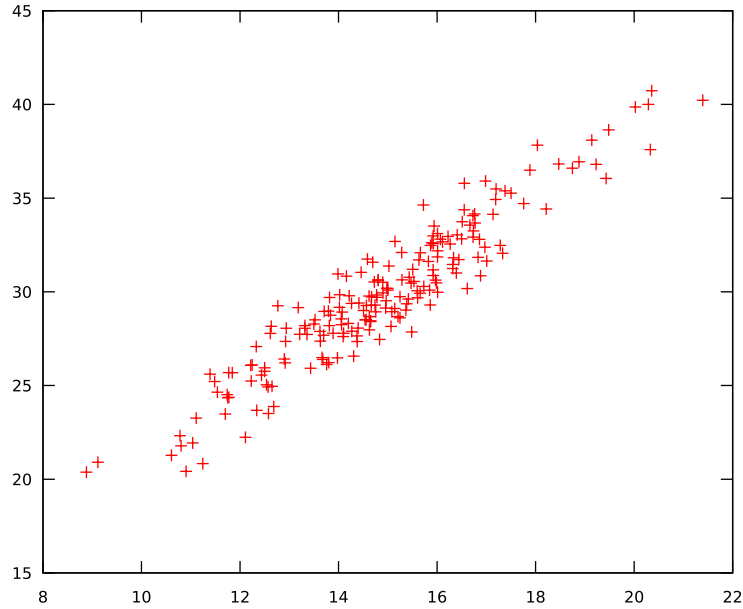}
\caption{Kaksigeenisten yksilöiden populaatiodiagrammi, jossa yksilöt esitetään
punaisina pisteinä.}
\label{scatterplot}
\end{figure}

Yksilöitä ja populaatioita voidaan havainnollistaa kuvan \ref{scatterplot} kaltaisilla
diagrammeilla. Differentiaalievoluutio vaatii, että geenit ovat reaalilukuja.
Kuvassa \ref{scatterplot} yksilöä vastaa piste, jonka x-koordinaatti on sama kuin yksilön
ensimmäinen geeni ja y-koordinaatti sama kuin toinen geeni. Populaatio koostuu
useasta yksilöstä eli useasta pisteestä. Diagrammeilla havainnollistetaan vain
kaksigeenisiä yksilöitä, koska kaksiulotteisella paperilla ei voida järkevästi
havainnollistaa kolmatta geeniä. Differentiaalievoluutio toimii tietysti
riippumatta geenien lukumäärästä. Jos geenejä kuitenkin on paljon, optimointi
kestää kauan.

Mutaatiossa kaikkia yksilön geenejä muutetaan satunnaisesti. Jos mutatoitava
yksilö $(x_i, y_i)$ tulkitaan vektoriksi $\beforemut$, mutaatio voidaan tulkita siten, että yksilöön
lisätään satunnainen mutaatiovektori $\mutvect$ josta saadaan mutatoitu vektori $\aftermut$ eli
\begin{eqnarray}
\aftermut & = & \beforemut + \mutvect \label{mutation}.
\end{eqnarray}

\begin{figure}[tb]
\igraph{0.7}{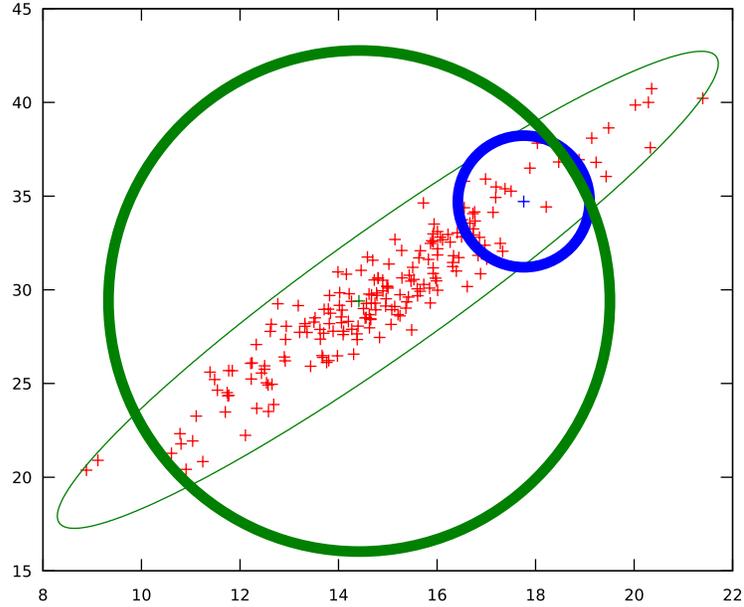}
\caption{Populaation yksilöiden jakauma on vihreän ellipsin muotoinen. Mutaation
pitäisi pystyä tuottamaan mikä tahansa piste tämän ellipsin sisäpuolelta, mutta
ei ulkopuolelta. Hyperpallomutaatio ei sovi kuvan populaatioon, sillä sininen
ympyrä on isoakselin suunnassa liian pieni, mutta vihreä
ympyrä on pikkuakselin suunnassa liian suuri.}
\label{hyperball}
\end{figure}

Mutaatiovektoria $\mutvect$ ei saa tietenkään valita miten tahansa. Kuvassa
\ref{hyperball} havainnollistetaan asiaa. Koska populaation yksilöt ovat asettuneet
likimain ellipsin muotoiselle alueelle, kelpoisuusfunktio on sellainen, että tällä
alueella on tietyssä mielessä hyviä yksilöitä. Uusia yksilöitä pitäisi siis
etsiä tämän ellipsin sisäpuolelta tai läheltä. Lisäksi on edullista, että
mutaatio tekee riittävän suuria muutoksia joka suunnassa.

\begin{figure}[tb]
\igraph{1.0}{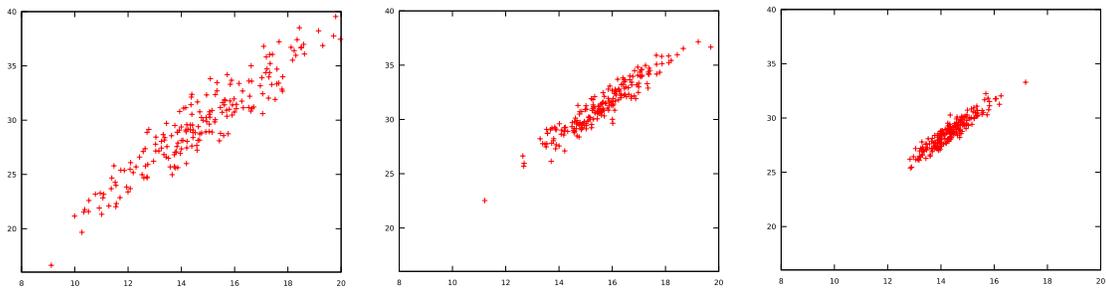}
\caption{Optimointi siirtää yksilöitä lähemmäs optimia, joten populaatiodiversiteetti pienenee optimoinnin aikana.}
\label{convergence}
\end{figure}

Mutaatiovektori voitaisiin valita satunnaisesti esimerkiksi $m$-ulotteisen
hyperpallon sisältä. Tässä tapauksessa hyperpallo olisi ympyrä. Havaitaan
kuitenkin, että hyperpallovalinta ei sovi kuvan populaatioon. Jos hyperpallon
säde on pieni, osa ellipsistä jää mutaatiomahdollisuuksien ulkopuolelle. Tämä
hidastaa konvergenssia, koska hyvä yksilö voi jäädä löytymättä. Jos taas
hyperpallon säde on suuri, riittävän suuri osa ellipsiä on
mutaatiomahdollisuuksien sisäpuolella, mutta mutaatio yrittää etsiä hyviä
yksilöitä myös ellipsin ulkopuolelta. Tämäkin hidastaa konvergenssia, koska
mutaatio löytää paljon huonoja yksilöitä. Mutaatiovektori pitäisi siis valita
sellaisen kuvion sisältä, joka on samanmuotoinen kuin yksilöiden jakauma.
Tällöin koko hyvien parametrien alue etsitään mahdollisimman tehokkaasti
jokaisessa suunnassa. Lisäksi populaation konvergoituessa kohti minimiä
tarvittavien mutaatioiden voimakkuus pienenee (kuva \ref{convergence}). Mutaatioiden suuruusluokan
on siis muututtava samalla, kun populaatiodiversiteetin suuruusluokka muuttuu. Yksi monimutkainen tapa valita sopiva
satunnainen mutaatiovektori on laskea
tilastollisia tunnuslukuja populaatiosta ja muodostaa niiden avulla jakauma,
joka muistuttaa populaation yksilöjakaumaa. Esimerkiksi jakaumaksi voitaisiin valita $m$-ulotteinen
normaalijakauma, jonka kovarianssimatriisi ja keskiarvovektori laskettaisiin populaation
kovarianssista ja keskiarvosta. Se on kuitenkin hidasta ja vaikeaa.

Differentiaalievoluution merkittävin ero muihin evoluutiopohjaisiin
algoritmeihin on, että mutaatiovektori lasketaan valitsemalla populaatiosta kaksi satunnaista
yksilövektoria $\randfirst$ ja $\randsecond$ ja laskemalla niiden vektorierotus
$\randfirst - \randsecond$. Mutaatiovektori $\mutvect$ saadaan skaalaamalla tämä
vektorierotus, eli
\begin{eqnarray}
\mutvect & = & k_m(\randfirst - \randsecond), \label{vectdiff}
\end{eqnarray}
missä $k_m$ on skaalauskerroin eli mutaatiovakio. Röntgendiffraktiodataa
on aiemmin sovitettu\cite{Wormington} arvolla $k_m = 0{,}7$, joten tätä
arvoa käytetään tässä työssä toteutetussa optimointialgoritmissa. Vektorierotus
on jokaisessa suunnassa suunnilleen samaa suuruusluokkaa kuin populaatiodiversiteetti
tässä suunnassa.

Risteytys yksilöille $\vect{u_1}$ ja $\vect{u_2}$ tehdään valitsemalla jokainen
geeni erikseen satunnaisesti joko yksilöstä $\vect{u_1}$ tai $\vect{u_2}$.
Todennäköisyys, että tietty geeni valitaan yksilöstä $\vect{u_2}$ on $k_c \in
[0,1]$. Jos siis $k_c$ on pieni, painotetaan yksilöä $\vect{u_1}$, ja jos se on
suuri, painotetaan yksilöä $\vect{u_2}$. Yksilöitä painotetaan symmetrisesti
kun $k_c = 0{,}5$. Tässä työssä käytetään yksinkertaisuuden vuoksi arvoa $k_c =
0{,}5$. Alkuperäinen DE käytti erilaista risteytystä\cite{DE}, mutta DE:n tekijät
toteuttivat myöhemmin myös tässä käytetyn risteytysmenetelmän\cite{DEbin},
jossa tosin varmistettiin, että vähintään yksi geeni tulee yksilöstä
$\vect{u_2}$. Ohjelmointiteknisistä syistä tämä merkitykseltään vähäinen
tarkistus jätettiin tekemättä tässä työssä, sillä se monimutkaistaa ja hidastaa
risteytyksen vektorisoitua Matlab-ohjelmakoodia, jossa ehtolauseiden määrä
halutaan pitää niin pienenä kuin mahdollista.

\begin{figure}[tb]
\igraph{0.7}{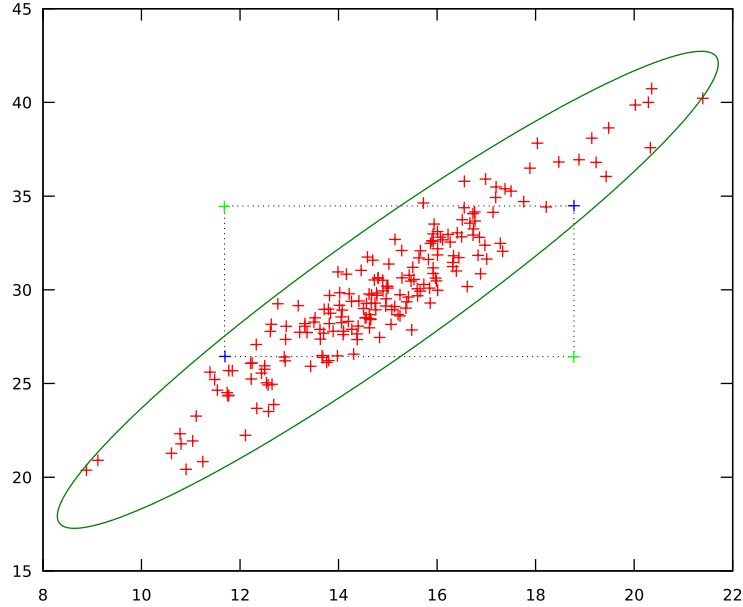}
\caption{Kahden sinisen yksilön risteytys alkuperäisessä koordinaatistossa tuottaa
joko jommankumman alkuperäisen yksilön tai jommankumman vihreän yksilön, jotka
ovat selvästi populaatiojakauman ulkopuolella.}
\label{crossover}
\end{figure}

Tietyn muotoisilla yksilöjakaumilla risteytystä vaivaa sama ongelma kuin
yksinkertaista hyperpallomutaatiota. Kuvan \ref{crossover} mukaisella yksilöjakaumalla
risteytys joko luo uuden yksilön, joka on selvästi yksilöjakauman ulkopuolella,
tai sitten risteytys antaa jommankumman alkuperäisistä yksilöistä. Risteytys ei
siis toimi niin kuin sen on tarkoitus toimia, eli luoda yksilö, joka muistuttaa
muita populaation yksilöitä, mutta jonka geenit eivät ole tarkasti samat kuin
millään muulla yksilöllä. Tässä tapauksessa geenien välillä on
vuorovaikutuksia, jonka seurauksena yhden geenin muuttaminen heikentää selvästi
yksilön kelpoisuutta, ellei toista geeniä muuteta samalla. Ongelma johtuu
lähinnä geenien koodaustavasta. Koordinaatistokierrolla saadaan uusi
koordinaatisto, jossa geenien välillä ei enää ole vuorovaikutuksia, ja
risteytys tässä koordinaatistossa toimii niin kuin risteytyksen pitääkin
toimia.

Röntgenheijastusdataan sovituksessa on aiemmin havaittu tällaisia geenienvälisiä
vuorovaikutuksia, jotka on ratkaistu monimutkaisemmalla optimointialgoritmilla,
jossa käytetään pääkomponenttianalyysiin (PCA, \emph{principal component analysis}) perustuvaa
koordinaatistorotaatiota\cite{CovGA}, joka on myöhemmin yleistetty myös
tehokkaammalla mutta hitaammalla ja monimutkaisemmalla riippumattomien
komponenttien analyysilla (ICA, \emph{independent component analysis}) \cite{ICAGA}. Koska ongelma on lähinnä huonosti
optimoitavaan funktioon soveltuvassa risteytysoperaattorissa, eikä siinä
optimointialgoritmissa, jossa risteytysoperaattoria käytetään,
koordinaatistorotaatio sopii myös käytettäväksi differentiaalievoluution
risteytysoperaattorissa. Tässä käytetään yksinkertaisuuden vuoksi
pääkomponenttianalyysia.

Kierretty koordinaatisto valitaan siten, että kovarianssimatriisi on
diagonaalimatriisi. Tällöin varsinaista kovarianssia eli geenienvälistä
lineaarista riippuvuutta ei ole, koska kovarianssimatriisin diagonaalielementit
kertovat yksittäisen geenin varianssin populaatiossa. Jos $n$-geenisiä
yksilöitä esitetään $1 \times n$-vaakavektoreina, ja $m$-yksilöinen populaatio
esitetään $m \times n$-matriisina $\vect{P}$ ja oletetaan, että jokaisen
geenin keskiarvo populaatiossa on 0, kovarianssimatriisi määritellään
\begin{equation}
\vect{C}  =  \mathrm{Cov}(\vect{P})
          =  \frac{1}{m-1}\trans{\vect{P}}\vect{P}, \label{Cov}
\end{equation}
ja on määritelmänsä perusteella symmetrinen ($\trans{\vect{C}} = \vect{C}$). Jos jonkin geenin
keskiarvo populaatiossa ei ole 0, yhtälössä (\ref{Cov}) esiintyvän populaatiomatriisin
$\vect{P}$ geeneistä on vähennettävä niiden populaatiokeskiarvo, jolloin
populaatiokeskiarvoiksi saadaan 0.

Jos nyt koordinaatistoa kierretään rotaatiomatriisilla $\vect{T}$, uusi
populaatio on $\vect{P}\vect{T}$. Tässä populaatio kerrotaan rotaatiomatriisilla
oikealta, koska yksilöt ovat vaakavektoreita. Uusi kovarianssimatriisi, jonka halutaan olevan
yhtäsuuri kuin eräs diagonaalimatriisi $\vect{\Lambda}$, on
\begin{equation}
\mathrm{Cov}(\vect{P}\vect{T}) = \frac{1}{m-1}\trans{\vect{T}}\trans{\vect{P}}\vect{P}\vect{T}
                               = \trans{\vect{T}}\vect{C}\vect{T}
                               = \vect{\Lambda} \label{covrot}.
\end{equation}

Koska $\vect{T}$ on rotaatiomatriisi ($\vect{T}^{-1} = \trans{\vect{T}})$, yhtälö
(\ref{covrot}) saadaan muotoon
\begin{eqnarray}
\vect{C}\vect{T} & = & \vect{T}\vect{\Lambda},
\end{eqnarray}
mikä on matriisin $\vect{C}$ ominaisarvo-ongelma. Symmetrisellä
reaalimatriisilla ominaisarvot ovat reaalisia, ja niitä vastaavat reaaliset
ominaisvektorit ovat ortogonaaliset ja ne voidaan normalisoida. On siis
olemassa koordinaatisto, jossa kovarianssimatriisi saadaan diagonaaliseksi.
Tarvittavan rotaatiomatriisin $\vect{T}$ sarakkeet ovat normalisoituja
matriisin $\vect{C}$ ominaisvektoreita. Kuvattua menetelmää kutsutaan
pääkomponenttianalyysiksi.

\begin{figure}[tb]
\igraph{0.7}{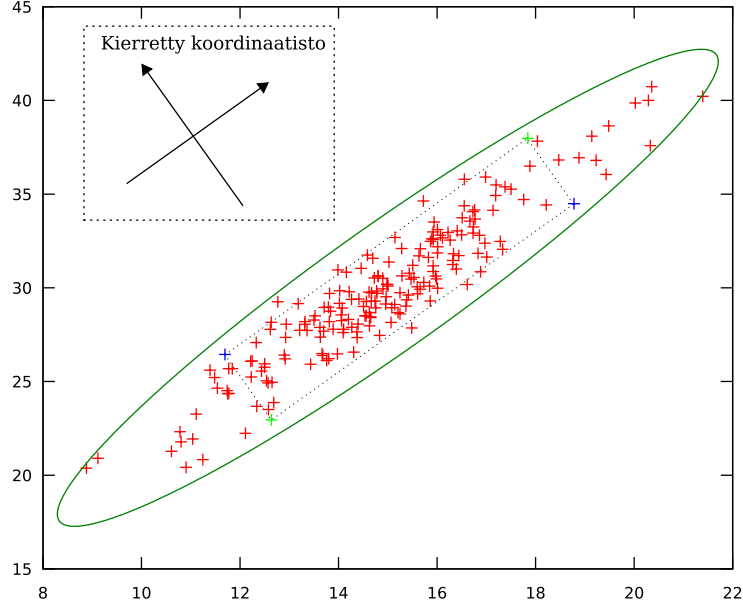}
\caption{Kahden sinisen yksilön risteytys pääkomponenttianalyysilla kierretyssä koordinaatistossa tuottaa
vain populaatiojakauman sisäpuolella olevia yksilöitä.}
\label{crossover-rot}
\end{figure}

Risteytys tehdään siten, että risteytettäviä yksilöitä $\vect{u_1}$ ja
$\vect{u_2}$ kierretään koordinaattirotaatiomatriisilla $\vect{T}$, jolloin
kierretyt yksilöt ovat $\vect{u_1}\vect{T}$ ja $\vect{u_2}\vect{T}$. Tässä
risteytettävät yksilöt on esitettävä vaakavektoreina, sillä muuten
matriisituloja ei ole määritelty. Risteytys tehdään näille kierretyille
yksilöille, jonka jälkeen risteytetty yksilö kierretään takaisin
käänteistoraatiomatriisilla $\vect{T}^{-1} = \trans{\vect{T}}$. Kuvasta \ref{crossover-rot}
havaitaan, että PCA-risteytytetyt yksilöt sijaitsevat populaatiojakauman
sisällä, mikä osoittaa että pääkomponenttianalyysin käyttö risteytyksessä toimii.

Differentiaalievoluutiota käytetään luomalla ensin satunnainen populaatio, ja
sitten suorittamalla iteraatiovaihetta niin monta kertaa, että populaatio on
konvergoitunut minimiin. Jokaiselle geenille $x_j$ määritellään rajat
$[x_{j,\mathrm{min}}, x_{j,\mathrm{max}}]$. Rajat voivat olla joko
fysikaalisia (esimerkiksi kerroksen paksuus ei voi olla nollaa pienempi) tai
voivat perustua aiempaan tietoon optimoitavasta ongelmasta. Satunnainen
populaatio luodaan valitsemalla jokaisen yksilön jokainen geeni satunnaisesti
sallittujen rajojen sisältä.

\begin{figure}[tb]
\igraph{1.0}{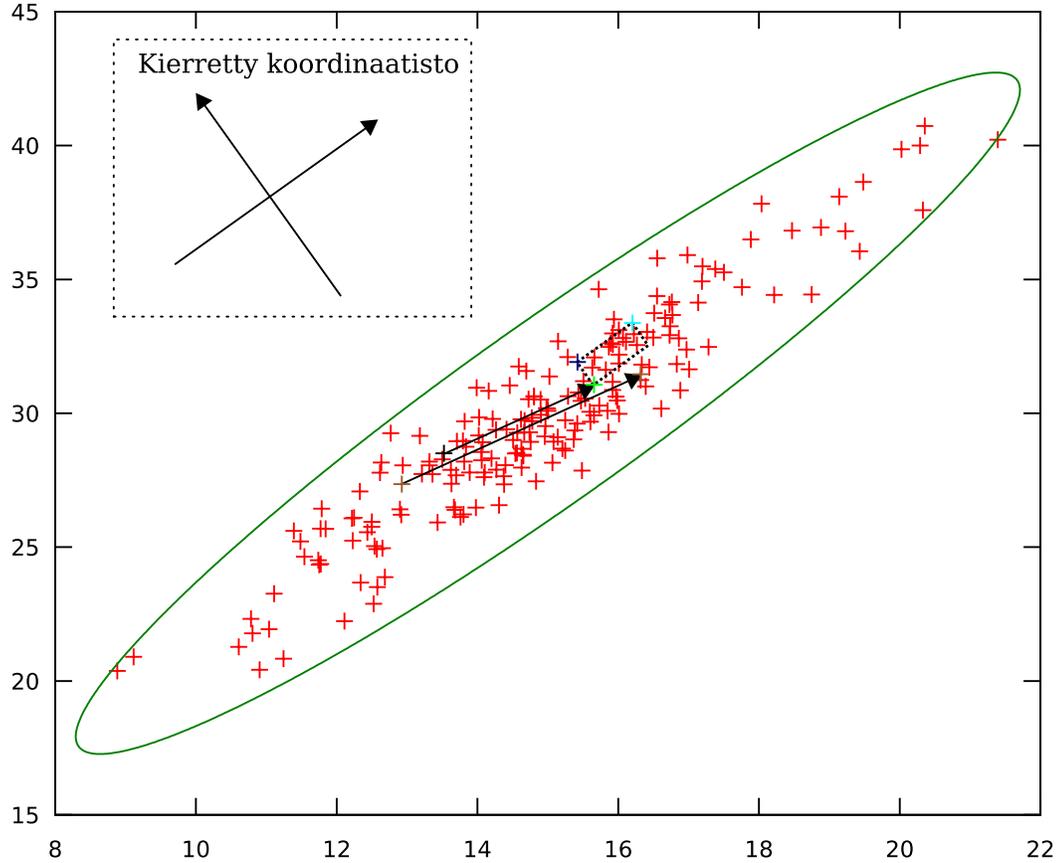}
\caption{Differentiaalievoluution iteraatio käsittelee populaation jokaista yksilöä.
Tässä kuvataan, miten differentiaalievoluutio muuttaa populaation ensimmäistä, vaaleansinistä
yksilöä. Populaatiosta valitaan kaksi satunnaista, ruskeaa yksilöä. Näiden skaalattu
vektorierotus lisätään parhaaseen, mustaan yksilöön jolloin saadaaan vihreä yksilö.
Alkuperäinen, vaaleansininen yksilö ja vihreä yksilö risteytetään, jolloin saadaan uusi,
tummansininen yksilö. Jos tummansininen yksilö on parempi kuin vaaleansininen, se korvaa
vaaleansinisen.}
\label{DEdiagram}
\end{figure}

Differentiaalievoluutioalgoritmin yksi iteraatio on mutaatioiden, risteytysten
ja valintojen yhdistelmä. Algoritmin käytössä on populaatio $\popvect,\ i \in 1..N$,
jonka parasta yksilöä merkitään $\bestvect$.
Jotkin versiot differentiaalievoluutiosta käyttävät mutaatiota vain parhaaseen
yksilöön, ja toiset käyttävät mutaatiota populaatiosta satunnaisesti valittuihin yksilöihin. Yleisesti
kaikkien DE-versioiden mutatoitava yksilö on
\begin{eqnarray}
\beforemut = \randthird + (\bestvect - \randthird)k_\lambda,
\end{eqnarray}
jossa $k_\lambda$ on siirrosvakio, $\randthird$ on populaatiosta satunnaisesti
valittu yksilö ja $\bestvect$ on populaation paras yksilö. Siirrosvakion
$k_\lambda$ arvolla 1 käytetään vain parasta yksilöä, ja arvolla 0 käytetään
vain satunnaista yksilöä. Tässä työssä valittiin $k_\lambda = 1$. Arvo voi tietysti olla näiden välissä, jolloin
satunnaisesti valittua yksilöä siirretään parhaan yksilön suuntaan. Näitä mutatoitavia
yksilöitä lasketaan yhtä monta kuin populaatiossa on yksilöitä, joten
$i$ saa kokonaislukuarvot $1..N$. Mutatoitavat yksilöt mutatoidaan kaavalla (\ref{mutation}),
jonka skaalattu vektorierotus $\mutvect$ lasketaan erikseen jokaiselle mutatoitavalle
yksilölle kaavalla (\ref{vectdiff}). Mutatoitu yksilö $\aftermut$ risteytetään
kierretyssä koordinaatistossa populaation $i$. yksilön $\popvect$ kanssa,
jolloin saadaan jokaista populaatioyksilöä $\popvect$ kohden uusi risteytetty yksilö $\newvect$.
Kun kierretty koordinaatisto lasketaan pääkomponenttianalyysilla, on otettava
huomioon liukulukulaskennan tarkkuus. Rajallinen tarkkuus on ongelmallinen,
koska esimerkiksi kerroksen paksuuden SI-järjestelmässä kertova geeni on
suuruusluokkaa $10^{-9}$, mutta kerroksen koostumuksen kertova geeni on
suuruusluokkaa $1$. Pääkomponenttianalyysin ajaksi geenit normalisoidaan
sallittujen rajojen $[x_{j,\mathrm{min}}, x_{j,\mathrm{max}}]$
avulla samaan suuruusluokkaan.
Koska mutaatio saattaa siirtää mutatoidun yksilön jonkin geenin sallittujen rajojen ulkopuolelle,
myös yksilöt $\newvect$ saattavat sisältää geenejä rajojen ulkopuolelta. Tämän takia vielä
tarkistetaan, että kaikki geenit ovat sallittujen rajojen sisäpuolella. Jos jokin geeni
ei ole näiden rajojen sisäpuolella, se korvataan valitsemalla sille arvo satunnaisesti
rajojen sisältä. Jokaisen populaation yksilön $\popvect$ kelpoisuutta verrataan yksilön $\newvect$ kelpoisuuteen,
ja näistä yksilöistä parempi jää populaatioon seuraavalle iteraatiolle.

\subsection{Sovitusohjelmat \label{software}}

Edellisissä luvuissa esitetyn teorian pohjalta kirjoitettiin tietokoneohjelma,
joka pystyy laskemaan teoreettisia röntgendiffraktiokuvaajia ja sovittamaan
näitä kuvaajia mittaustuloksiin differentiaalievoluutiolla. Tietokoneohjelma
kirjoitettiin Javalla ja Matlabilla. Ohjelman käyttöliittymä perustuu aiemmin
kirjoitettuun röntgenheijastusohjelmaan. Käyttöliittymä on kehitetty käyttäjien
palautteen perusteella mahdollisimman tehokkaaksi.

Matlab-osuus sisältää täydellisen röntgendiffraktiokirjaston, joka pystyy
laskemaan diffraktiointensiteetin epitaktisille kerrosrakenteille. Kirjasto on
tehty mahdollisuuksien mukaan yleiskäyttöiseksi, joten sitä voi hyödyntää
muutenkin röntgendiffraktiotutkimuksen apuna. Matlabilla toteutettiin myös
yleiskäyttöinen differentiaalievoluutiokirjasto optimointiongelmien
ratkaisuun.  DE-kirjasto ei ole mitenkään sidottu röntgendiffraktioon, vaan
sitä voi käyttää muihinkin optimointiongelmiin. DE-kirjastosta pyrittiin
tekemään mahdollisimman nopea, mikä Matlabin kaltaisella ohjelmointikielellä
edellyttää lähinnä yksinkertaisuutta. DE-algoritmin olennainen osuus on noin 40
koodiriviä pitkä. Lisäksi Matlab-osuus sisältää kaiken tarvittavan DE:n
yhdistämiseen röntgendiffraktiokäyrien sovitukseen, eli kelpoisuusfunktion ja
apufunktioita, jotka tekevät röntgendiffraktiokäyrien sovituksesta
yksinkertaisempaa. Matlab-koodi toimii myös ilmaisella Matlab-kloonilla, GNU
Octavella.

Tietokoneohjelmien käyttäjälle näkyvä toiminnallisuus tehtiin Javalla. Java
valittiin käyttöliittymäkieleksi, koska se on alustariippumaton. Javalle on
myös saatavilla erinomainen käyttöliittymäkirjasto Swing. Matlab valittiin
röntgendiffraktion ja optimointialgoritmin toteutuskieleksi, koska Matlabilla
on helppo kirjoittaa paljon matemaattisia laskutoimituksia sisältäviä
algoritmeja. Yleiskäyttöisenä ohjelmointikielenä Matlab on riittämätön.
Röntgendiffraktiokoodi jouduttiin kirjoittamaan Matlabin lisäksi myös Javalla,
koska Matlab-koodin integrointi muihin ohjelmointikieliin on kömpelöä. Tämä
johtuu siitä, että kommunikointi Matlab-tulkin kanssa tapahtuu tekstipohjaisten
virtojen kautta, eikä Matlab esimerkiksi tue rinnakkaisohjelmointia. 

Saman koodin toteutus kahteen kertaan aiheuttaa ylimääräistä painolastia
ylläpidolle. Javan huono soveltuvuus numeeriseen laskentaan aiheuttaa myös
ylimääräistä painolastia. Koko ohjelma olisi luultavasti kannattanut ohjelmoida
sellaisella kielellä, joka sopii hyvin sekä numeeriseen laskentaan että
yleiskäyttöiseen ohjelmointiin. C++ olisi ollut hyvä valinta, mutta hyviä
C++-käyttöliittymäkirjastoja ei ole saatavilla, ja C++ vaatii ohjelman
kääntämisen lähdekoodista erikseen eri alustoille. Tietysti käyttöliittymän
olisi voinut toteuttaa Javalla ja kaiken muun C++:lla, koska C++-koodin voi
integroida Javaan huomattavasti paremmin kuin Matlab-koodin.

Ohjelman käyttöliittymä koostuu kolmesta eri välilehdestä. Nämä ovat
kerrosmallieditori, manuaalinen sovitus ja automaattinen sovitus.
Kerrosmallieditorilla muodostetaan malli, joka sisältää eri materiaaleista
koostuvia ohutkerroksia päällekkäin. Välilehdellä määritellään jokaiselle
kerrokselle kaikki ominaisuudet, joista tärkeimpiä ovat materiaalit, joista
kerros koostuu. Manuaalisen sovituksen välilehdellä voi hienosäätää kerroksen
sovitettavia parametreja ja katsoa, kuinka ne vaikuttavat
diffraktiointensiteetin kuvaajaan. Kun kerrosmallin kaikki ominaisuudet ovat
säädetty kuntoon, automaattisen sovituksen välilehdellä voi etsiä
differentiaalievoluutiolla sovitusparametreille optimaaliset arvot.

\begin{figure}[t]
\begin{center}
\includegraphics[height=9cm]{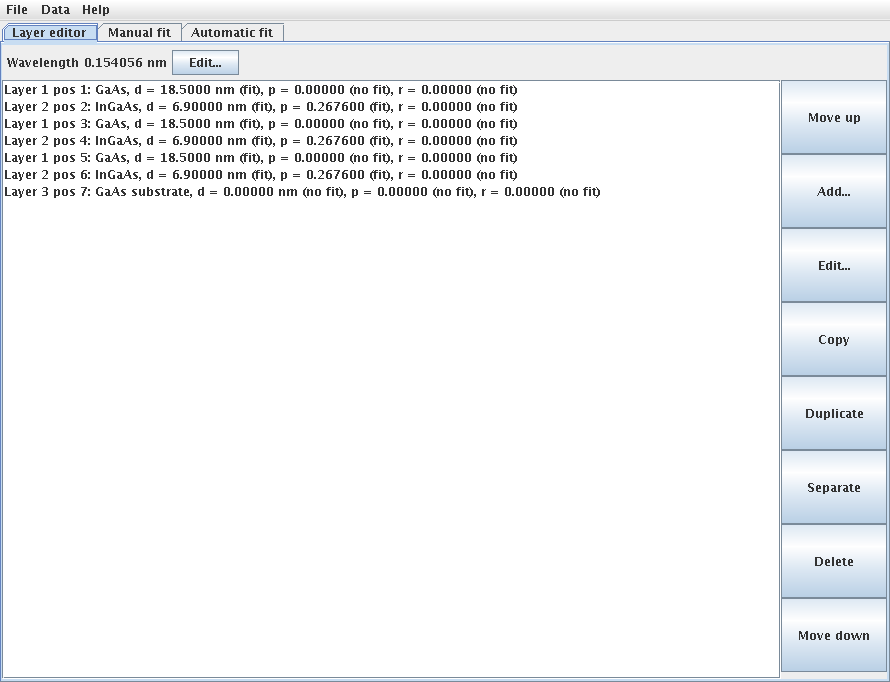}
\end{center}
\caption{Ohjelman kerrosmallieditori.}
\label{layered}
\end{figure}

Kuvassa \ref{layered} on välilehti, jolla rakennetaan kerrosmalli. Vasemmalla
olevassa listassa ovat kaikki kerrosmalliin kuuluvat kerrokset päällekkäin.
Pohjimmaisin kerros on substraatti, ja sen päällä on ohutkerroksia. Kerroksille
määritellään nimi, materiaalit, ja sovitusparametrit eli paksuus, koostumus ja
relaksaatio. Listassa ei näy materiaaleja, joten yleensä kerroksen nimessä
pitäisi kuvata lyhyesti materiaalit, joista kerros koostuu. Sama looginen
kerros voi toistua mallissa useammassa eri fyysisessä paikassa. Tällöin näiden
eri paikoissa olevien fyysisten kerrosten ominaisuudet sovitetaan yhdessä.

Esimerkiksi kuvan \ref{layered} mallissa kerroksia ovat GaAs-substraatti,
InGaAs ja GaAs. Näistä InGaAs ja GaAs toistuvat jaksollisesti kolmen periodin
verran, ja niiden paksuudet ja koostumukset sovitetaan yhdessä. GaAs-kerroksen
ja substraatin koostumus ovat jo tiedossa, joten niitä ei soviteta. Substraatin
paksuus oletetaan aina äärettömäksi, joten sitäkään ei soviteta. Kaikkien
kerrosten relaksaatio on määritelty nollaksi. Sovitettavia kerrosten
ominaisuuksia ovat siis GaAs- ja InGaAs-paksuudet ja InGaAs-koostumus, eli
niitä on kolme.

\begin{figure}[t]
\begin{center}
\includegraphics[height=9cm]{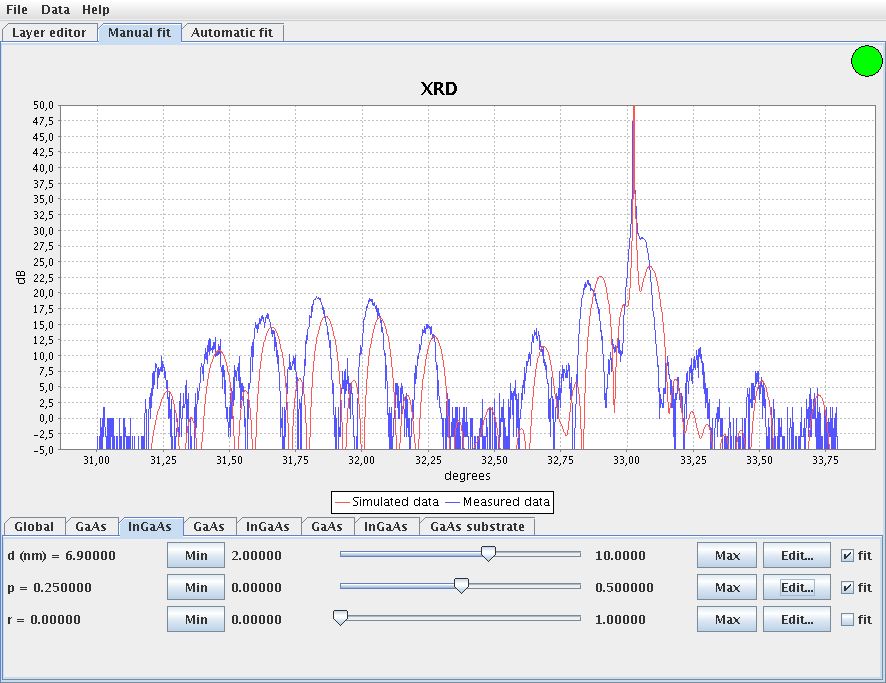}
\end{center}
\caption{Ohjelman manuaalisen sovituksen välilehti.}
\label{manfit}
\end{figure}

Kuvassa \ref{manfit} on sovitusohjelman välilehti, jolla kerrosmallin
ominaisuuksia voidaan hienosäätää automaattista sovitusta varten. Ylhäällä
olevaan kuvaajaan on piirretty simuloitu ja mitattu intensiteetti.
Pystyakselilla on ilmoitettu fotonimäärä sekunnissa desibeleinä. Fotonimäärä on
verrannollinen intensiteettiin, joka on tehosuure. 10 desibelin lisäys siis
vastaa intensiteetin kymmenkertaistumista. 0 dB vastaa yhtä fotonia sekunnissa.
Mittausdatassa näkyy voimakasta fotonilaskentakohinaa lyhyestä mittausajasta
johtuen. Kuvasta havaitaan, että yhtä mittauspistettä kohti intensiteettiä on
mitattu laskemalla saapuvia fotoneja kahden sekunnin ajan, sillä mittausdata on
diskreettiä, ja siinä esiintyvät pienimmät fotonitasot ovat noin -3 dB, 0 dB ja
2 dB, jotka vastaavat 0,5, 1 ja 1,5 fotonia sekunnissa.

Kerrosmallin ominaisuuksia voidaan muuttaa alhaalla olevista säätimistä.
Kuvassa näkyy InGaAs-kerrokseen liittyvät liukusäätimet, joilla voidaan muuttaa
kerrospaksuutta, koostumusta ja relaksaatiota, ja määritellä näille rajat,
joiden sisäpuolelta optimaalisia arvoja etsitään. Arvot voi myös määritellä
kiinteiksi, jolloin optimointialgoritmi ei muuta niitä yrittäessään sovittaa
simuloitua käyrää mittausdataan. InGaAs-kerroksen relaksaatio on määritelty
nollaksi, eikä optimointialgoritmi muuta sitä. Kuvaajan simuloitu käyrä on
piirretty paksuuden arvolla 6,9 nm ja koostumusparametrin $p$ arvolla 0,25,
mutta optimointialgoritmi etsii näiden arvoja väleiltä [2 nm, 10 nm] ja [0;
0,5].

\begin{figure}[t]
\begin{center}
\includegraphics[height=9cm]{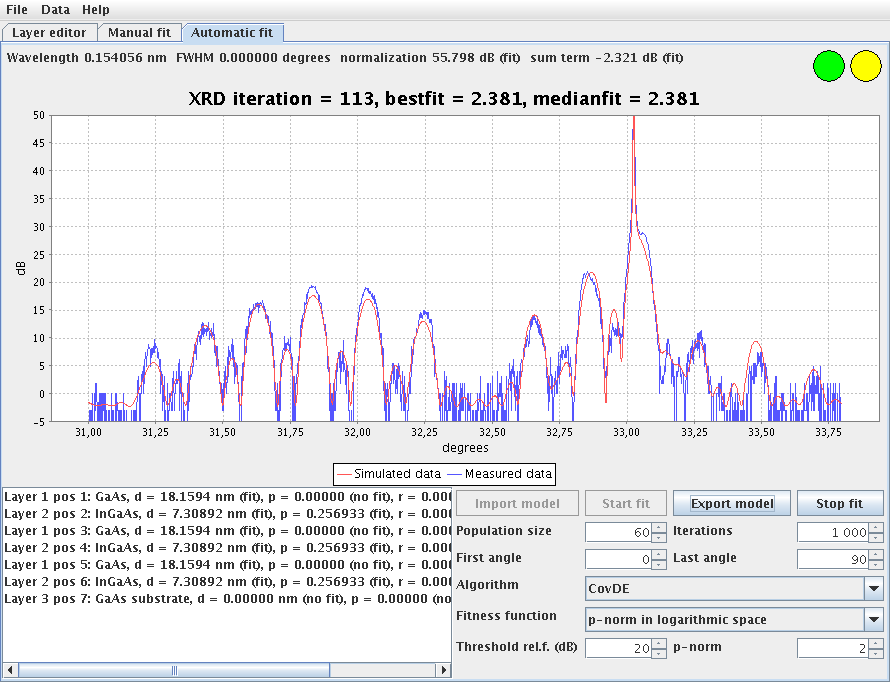}
\end{center}
\caption{Ohjelman automaattisen sovituksen välilehti.}
\label{autofit}
\end{figure}

Kuvassa \ref{autofit} on röntgendiffraktiosovitusohjelman automaattisen sovituksen
välilehti. Simuloitua käyrää sovitetaan juuri mittausdataan
differentiaalievoluutioalgoritmilla. Kaikkien GaAs-kerroksien paksuudet ja
In$_p$Ga$_{1-p}$As-kerrosten paksuudet ja koostumukset ovat lukittu samoiksi.
Populaation kooksi on määritelty 60, ja mutaatio-, risteytys- ja
valintavaiheita iteroidaan enintään 1000 kertaa. Iteraatioita on suoritettu
kuvassa vasta 113 kertaa, joiden jälkeen simuloitu käyrä vastaa jo hyvin
mittausdataa. Parhaimman yksilön kelpoisuus, 2,381, on sama kuin
mediaanikelpoisuus populaatiossa. Tämän voidaan tulkita tarkoittavan, että
populaatiodiversiteetti on hävinnyt, eikä optimointialgoritmi saa simuloitua
käyrää vastaamaan enää paremmin mittausdataa. Mitatut arvot näkyvät
kerrosmallilistasta, jossa GaAs-kerrosten paksuus on 18,2 nm,
In$_p$Ga$_{1-p}$As-kerrosten paksuus on 7,3 nm, ja In$_p$Ga$_{1-p}$As-kerrosten
koostumuksen kertova parametri $p = 0{,}257$.

\clearpage
\section{Tulokset}

\subsection{Simuloitujen käyrien vertailu \label{gidsl}}

Sovitusohjelman kriittisin osa on intensiteetin simulointi. Jos intensiteetti
lasketaan virheellisesti, ohjelma antaa vääriä mittaustuloksia. Jos taas
muualla tietokoneohjelmassa on virhe, se vaikuttaa lähinnä ohjelman
käytettävyyteen. Esimerkiksi virhe optimointialgoritmissa tekisi käyrän
sovituksesta hidasta tai mahdotonta. Tällöin virhe ei olisi piilossa, vaan
käyttäjälle olisi selvää, että mittaustuloksia ei saada ollenkaan.

Tämän takia uudella ohjelmalla simuloituja intensiteettikuvaajia verrattiin
Philipsin röntgendiffraktometrin ohjelmistolla simuloituihin kuvaajiin. Neljään
aiemmin julkaistuun mittaukseen\cite{GaAsN} liittyvät simulaatiot suoritettiin
uudestaan uudella ohjelmalla. Näytteet ovat hyvin yksinkertaisia, yksittäisiä
epitaktisia GaAs$_{1-p}$N$_{p}$-kerroksia GaAs-substraatin päällä.
GaAs$_{1-p}$N$_{p}$-kerrosten koostumusparametrit $p < 0{,}01$ ja paksuudet ovat
210\us{nm}--560\us{nm}. Kerrokset ovat täysin jännittyneitä. Substraatti, jonka todellinen paksuus on
$350\ \mathrm{\mu{}m}$, oletetaan äärettömän paksuksi. Molempien yhdisteiden
kiderakenne on sinkkivälkehila, ja näytteen pinta on likimain tason $(100)$
suuntainen. Kulma tason $(100)$ ja todellisen pinnan välillä on alle
$0{,}5\grad$. Diffraktio on mitattu käänteishilan vektorista $(400)$.

Uuden ohjelman materiaalitietokantaan asetettiin samat arvot kuin Philipsin
ohjelmassa, jotta tiedettäisiin, että mahdollinen poikkeama Philipsin ohjelman
ja uuden ohjelman välillä johtuu nimenomaan laskennan suorittavasta
ohjelmakoodista eikä pienistä eroista materiaalitietokannan lukuarvoissa.
Philipsin ohjelma käyttää GaAs:n hilavakiona arvoa $5{,}65368\ \textrm{Å}$ ja
sinkkivälkehilaan järjestäytyneen GaN:n hilavakiona arvoa $4{,}5034\ \textrm{Å}$.
Philipsin ohjelmassa Poissonin luvut ovat GaAs:lle 0,311 ja GaN:lle 0,33.
Suskeptiivisuuden laskentaan käytettäviä tietokantoja
\cite{Waasmaier,Henke,Peng} ei muutettu vastaamaan Philipsin ohjelman arvoja,
sillä suskeptiivisuus ei vaikuta yhtä paljon kuvaajiin kuin Poissonin luku
ja hilavakio.

\begin{figure}[p]
\igraph{0.7}{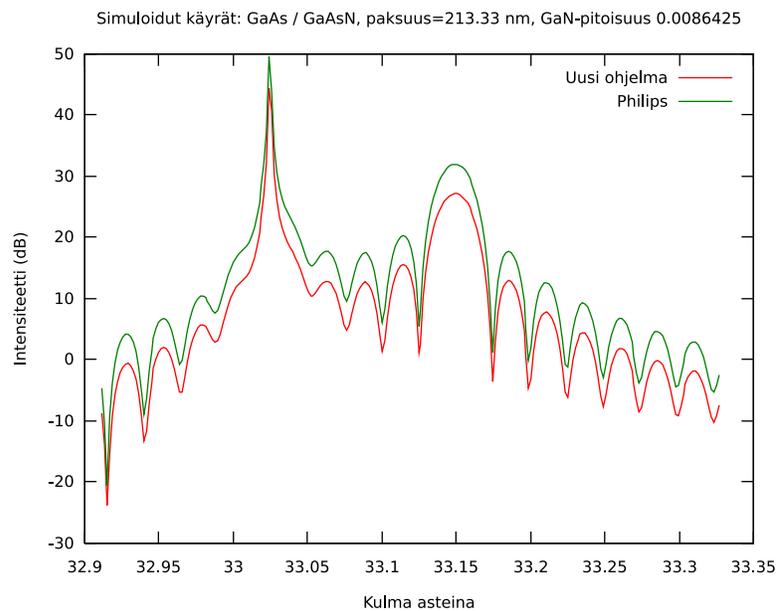}
\caption{Philipsin Smoothfit-ohjelmalla ja uudella ohjelmalla simuloidut intensiteettikuvaajat.
         Kuvaajat on siirretty selkeyden vuoksi pystysuunnassa eri korkeuksille.}
\label{philipscomp}
\end{figure}

Kuvassa \ref{philipscomp} ovat molemmilla ohjelmilla simuloidut kuvaajat
näytteelle A. Näytteen A kerroksen koostumus on GaAs$_{0{,}9913575}$N$_{0{,}0086425}$ ja paksuus
$213{,}33\us{nm}$. Näytteen todellisen pinnan poikkeamaa pinnasta $(100)$ ei ole
otettu tässä kuvaajassa huomioon. Molemmat kuvaajat ovat hyvin tarkalleen
samanlaiset. Myös näytteiden B--D kuvaajat (ei kuvassa) ovat molemmilla
ohjelmilla simuloituna samanlaiset. Vertailu osoittaa, että ohjelma toimii
ainakin yksikerroksisille rakenteille. Diffraktiointensiteetti monikerroksisista
rakenteista lasketaan samalla algoritmilla, mutta yhtä vaihetta vain
iteroidaan useamman kerran.

\begin{figure}[p]
\igraph{0.7}{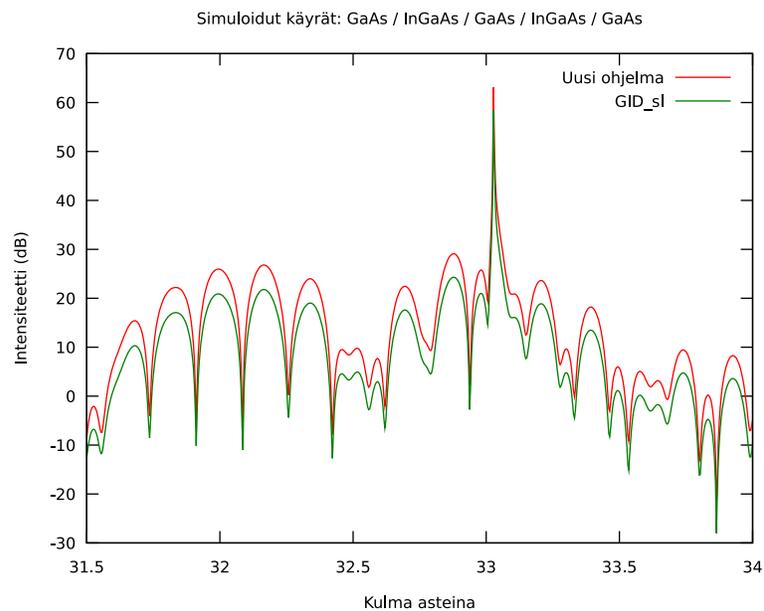}
\caption{GID\_sl-ohjelmalla ja uudella ohjelmalla simuloidut kuvaajat
         monikerrosrakenteelle. Kuvaajat on siirretty selkeyden vuoksi
         pystysuunnassa eri korkeuksille.}
\label{gidslcomp}
\end{figure}

Monikerrosrakenteiden simulaation toimivuutta testattiin vertailemalla
simuloituja kuvaajia GID\_sl-ohjelmalla\cite{gidsl} suoritettuihin
simulaatioihin. GID\_sl käyttää tässä työssä käytetyistä iteraatiokaavoista
poikkeavaa rekursiivista matriisialgoritmia\cite{RMA}.
Simulaatioissa käsitellään kuvitteellista näytettä, joka
sisältää GaAs-substraatin päällä In$_{0{,}2}$Ga$_{0{,}8}$As-, GaAs-,
In$_{0{,}2}$Ga$_{0{,}8}$As- ja GaAs-kerrokset (alimmasta ylimpään). GaAs-kerrosten
paksuus on $20\us{nm}$ ja InGaAs-kerrosten paksuus $10\us{nm}$. InGaAs-kerrokset
ovat täysin jännittyneitä. Pintana
on tarkalleen taso $(100)$. Diffraktio simuloitiin käänteishilan vektorille
$(400)$. GaAs:n ja InAs:n hilavakioina olivat molemmissa ohjelmissa
$5{,}6532\ \textrm{Å}$ ja $6{,}0584\ \textrm{Å}$. Poissonin vakiot GaAs:lle
ja InAs:lle olivat 0,3117 ja 0,352. Molemmilla ohjelmilla simuloidut
kuvaajat ovat kuvassa \ref{gidslcomp}. Kuvaajat ovat hyvin tarkalleen samanlaiset,
mikä osoittaa että diffraktion simulointi toimii myös monikerrosrakenteille.

\subsection{Esimerkkimittauksia \label{meas}}

Aiemmin huomattiin, että ohjelma laskee samanlaisen simuloidun käyrän kuin
muutkin tietokoneohjelmat, joten se saattaa toimia mittaustulosten
analysoinnissa. Käyränsovitusta ei ole kuitenkaan vielä tähän mennessä
testattu, joten käyränsovituskoodissa olevat ohjelmointivirheet voivat estää
ohjelman toiminnan. Jotta varmistutaan, että ohjelma todella toimii, sillä on
analysoitava uudestaan aiemmin suoritettuja mittauksia.

Luvussa \ref{gidsl} varmistettiin, että ohjelma laskee samanlaisen käyrän kuin
aiemmin tehtyihin mittauksiin liittyvät simulaatiot. Jos sovitusalgoritmi
toimii, sen olisi siis annettava samanlaiset mittaustulokset kuin aiemmin
julkaistut\cite{GaAsN}, Philipsin ohjelmalla määritetyt tulokset. Kaikki neljä
mittausta analysoitiin uudestaan, ja ohjelma antoi samanlaiset mittaustulokset.
Kaikissa mittauksissa GaAs-substraatin päällä oli GaAsN-kerros, jonka paksuus
ja GaN-pitoisuus oli määritettävä. Sovitettavia parametreja olivat myös
taustasäteilyn intensiteetti, röntgenlähteen intensiteetti ja
diffraktiovektorin ja pinnannormaalin välinen kulma $\phi$.
Materiaalitietokannan arvot (GaAs-hilavakio $5{,}65368\ \textrm{Å}$,
GaN-hilavakio $4{,}5034\ \textrm{Å}$, GaAs:n Poissonin luku $0{,}311$ ja GaN:n
Poissonin luku $0{,}33$) olivat vertailukelpoisuuden varmistamiseksi otettu
Philipsin ohjelmasta. Aiemmin julkaistut tulokset ja uudella ohjelmalla saadut tulokset
ovat koottu taulukkoon \ref{results}. Tulokset ilmoitetaan kahden desimaalin
tarkkuudella, koska ne julkaistiin aiemmin tällä tarkkuudella\cite{GaAsN}.

\begin{table}[h!]
\caption{Aiemmin julkaistujen\cite{GaAsN} Philipsin ohjelmalla analysoitujen mittausten
käsittely uudella ohjelmalla. Kaikki
mittaustulokset ovat kahden desimaalin tarkkuudella samat. Näytteiden A--B mittauksiin ei saatu
sovitettua käyrää hyvin, joten näytteet saattavat olla epäideaalisia. Näytteiden C--D mittauksiin
sovitus onnistui hyvin.}
\label{results}
\begin{center}
\setlength{\tabcolsep}{1mm}
\setlength{\extracolsep}{1mm}
\begin{tabular}{|l|l|l|l|l|l|} \hline
& Paksuus & GaN-pitoisuus & Paksuus & GaN-pitoisuus & Sovituksen \\
& (Philips) & (Philips) & (uusi ohjelma) & (uusi ohjelma) & laatu \\ \hline
Näyte A & 210 nm & 0,86\% & 210 nm & 0,86\% & huono \\ \hline
Näyte B & 480 nm & 0,85\% & 480 nm & 0,85\% & huono \\ \hline
Näyte C & 470 nm & 0,83\% & 470 nm & 0,83\% & hyvä \\ \hline
Näyte D & 560 nm & 0,86\% & 560 nm & 0,86\% & hyvä \\ \hline
\end{tabular}
\end{center}
\end{table}

\begin{figure}[p]
\igraph{0.7}{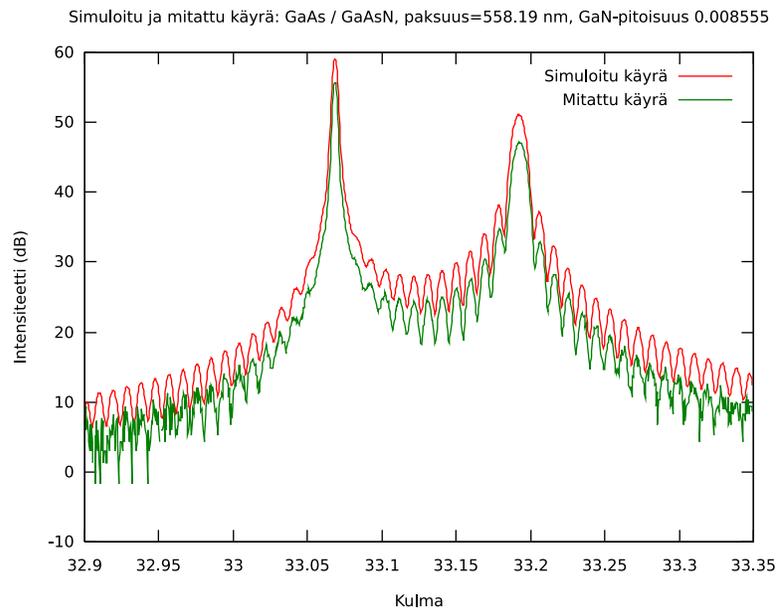}
\caption{Näytteen D mittaus ja siihen sovitettu käyrä. Simuloitu käyrä sopii hyvin mitattuun. Selkeyden vuoksi simuloitua käyrää on siirretty ylöspäin.}
\label{sample-d}
\end{figure}

\begin{figure}[p]
\igraph{0.7}{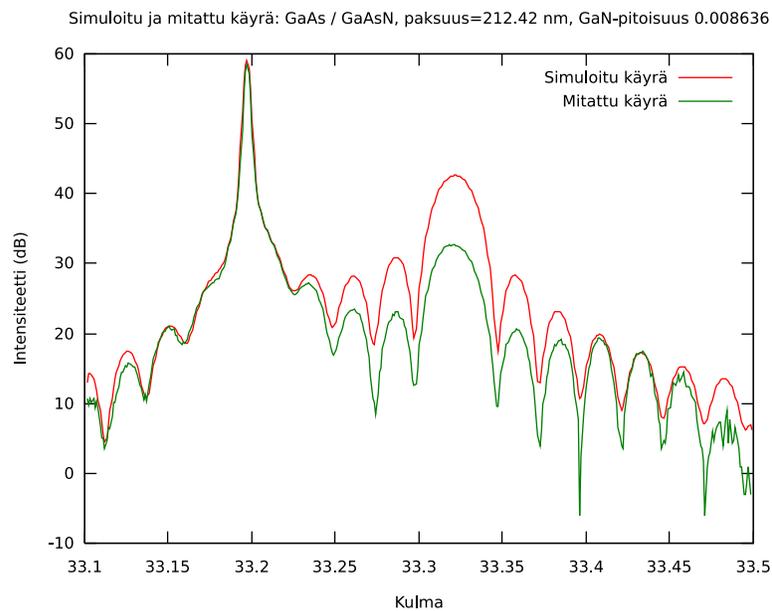}
\caption{Näytteen A mittaus ja siihen sovitettu käyrä. Simuloidun käyrän värähtelyjen jakso on sama kuin mittauksessa, mutta yksittäisten värähtelyjen voimakkuus on erilainen. Selkeyden vuoksi simuloitua käyrää on siirretty ylöspäin.}
\label{sample-a}
\end{figure}

Näytteistä A ja B tehtyjen mittaustulosten sovitus onnistui huonosti.
Simuloidun käyrän intensiteetin värähtely kulman funktiona saatiin vastaamaan
mittauksia, mutta värähtelyjen intensiteetin suuruudessa on selviä eroja.
Huono sovitus ei johdu uudesta tietokoneohjelmasta, sillä luvussa \ref{gidsl}
todettiin, että se laskee samanlaisia simuloituja käyriä kuin muutkin
tietokoneohjelmat. Selkeä ero mitatun ja simuloidun käyrän välillä on
havaittavissa myös aiemmin julkaistuissa kuvaajissa\cite{GaAsN}. Näytteet A ja
B eivät luultavasti ole kovin hyvälaatuisia, koska simuloitu käyrä ei vastaa
mitattua. Esimerkiksi ohutkerros ei välttämättä ole homogeeninen toisin kuin
luvussa \ref{xrd} oletettiin.

Simuloitu käyrä sopii näytteistä C ja D tehtyihiin mittauksiin huomattavasti
paremmin kuin näytteiden A ja B tapauksessa. Tämä ei kuitenkaan tarkoita, että
kerros on välttämättä aiemmin oletetun mukainen, eli homogeeninen ja täysin
jännittynyt.  Se tarkoittaa, että näytteiden mahdollista poikkeamaa
ideaalisuudesta ei havaita näissä mittauksissa. On esimerkiksi helppo
kokeilemalla todeta, että olettamalla kerros vain osittain jännittyneeksi
tietyllä relaksaatiolla $R$, saadaan simuloitu käyrä sopimaan mittaukseen aivan
yhtä hyvin, mutta GaN-pitoisuudelle saadaan tässä tapauksessa erilainen
mittaustulos. Tämä johtuu lähinnä siitä, että röntgendiffraktio on herkkä
pintaa vastaan kohtisuoralle hilavakiolle, mutta ei ole herkkä
suskeptiivisuudelle ja pinnansuuntaiselle hilavakiolle. Kohtisuora hilavakio
riippuu sekä relaksaatiosta että GaN-pitoisuudesta, eikä pelkkä kohtisuora
hilavakio tuntemalla voida ratkaista sekä relaksaatiota että GaN-pitoisuutta
yksikäsitteisesti.

Kuvassa \ref{sample-d} on näytteen D mittaus ja siihen sovitettu käyrä. Tähän
mittaukseen sovitus onnistui hyvin. Kuvan \ref{sample-a} näytteen A mittaukseen
käyrä ei sen sijaan sopinut hyvin.

Algoritmin soveltuvuutta vaikeisiin sovitusongelmiin testattiin luomalla
teoreettinen kerrosmalli, jossa on GaAs-substraatin päällä kuusi erilaista
kerrosta, ja sovittamalla intensiteettikäyrä tästä kerrosmallista laskettuun
teoreettiseen intensiteettikäyrään. Tässä kohdassa olisi voinut myös käyttää
teoreettisen intensiteettikäyrän sijasta oikeaa mittausta, mutta riittävän
korkealaatuista mittausta ei ollut saatavilla. Kerrosmallin jokaiselle
GaAs-kerrokselle asetettiin erilainen paksuus, ja jokaiselle
In$_p$Ga$_{1-p}$As-kerrokselle asetettiin erilainen koostumusparametri $p$ ja paksuus.
Kerrosmallin rakenne ja kerrosten koostumukset ja paksuudet ovat taulukossa
\ref{theorfit}. Diffraktiovektorin ja pinnannormaalin väliseksi kulmaksi
asetettiin $\phi = 0{,}03\grad$. Lisäksi malliin liittyy tulevan röntgensäteen
intensiteetti, joka asetettiin 70 dB:iin ja taustasäteilyn intensiteetti, joka
asetettiin 0 dB:iin. Oikeaan mittaukseen liittyvää epäideaalisuutta yritettiin
mallintaa pehmentämällä teoreettista käyrää konvolvoimalla se
puoliarvoleveydeltään 0,005\grad{} gaussisen jakauman kanssa ja mallintamalla
fotonilaskennasta syntyvää kohinaa Poissonin prosessilla, jossa yhden fotonin
intensiteetti oli 0 dB. Teoreettiseen käyrään laskettiin tuhat pistettä
tulokulmaväliltä 32--34\grad.

Tähän simuloituun teoreettiseen käyrään sovitettiin toinen simuloitu käyrä,
joka pehmennettiin myös 0,005\grad{} gaussisella konvoluutiolla. Sovitusparametreja oli 12, joista 6 oli
kerroksien paksuuksia, 3 kerroksien koostumuksia ja loput olivat tulevan säteen
intensiteetti, taustasäteilyn intensiteetti ja kulma $\phi$. Kerrospaksuuksien
ja kerrosten koostumusten arvot etsittiin taulukossa \ref{theorfit} mainituilta
väleiltä. Tulevan säteen intensiteetti ja taustasäteilyn intensiteetti
etsittiin väleiltä $55$--$85$ dB ja $-30$--$5$ dB. Kulma $\phi$ etsittiin
väleiltä $-0{,}05$--$0{,}05$\grad. Sovitusalgoritmissa käytettiin populaatiokokoa
150.

\begin{figure}[p!]
\igraph{1.0}{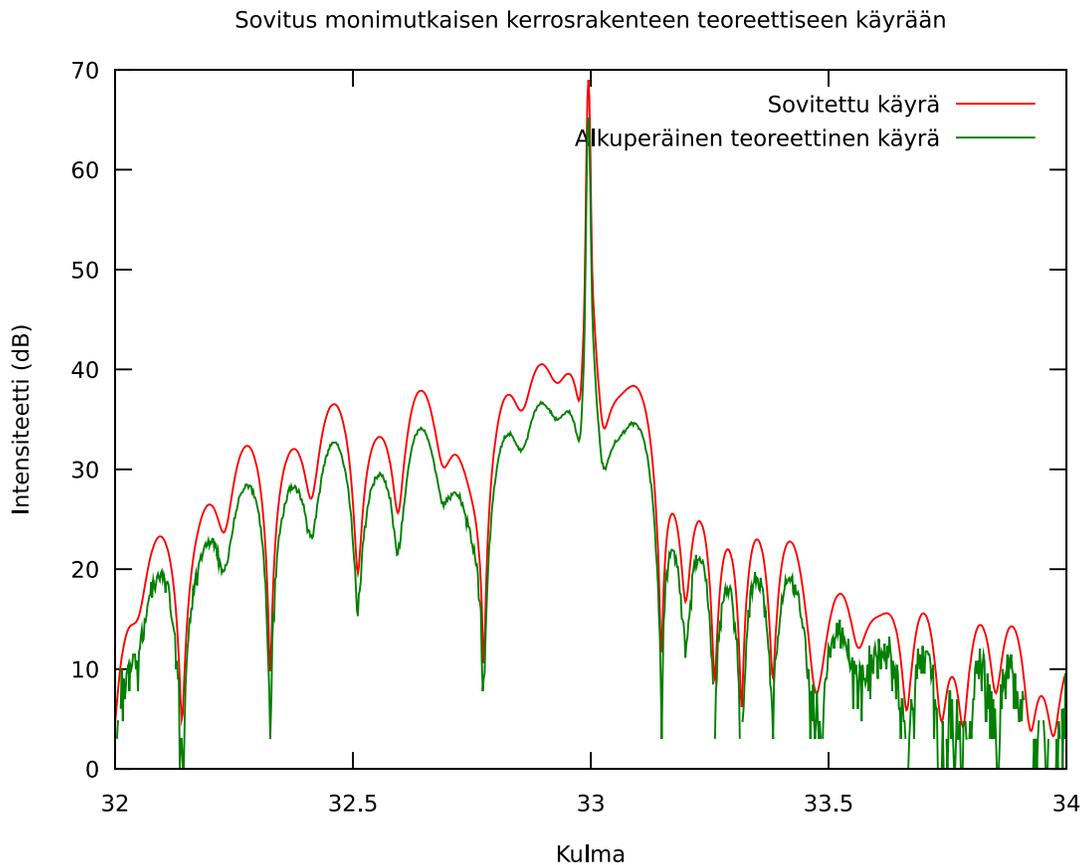}
\caption{Monimutkaisesta monikerrosrakenteesta laskettu teoreettinen
intensiteettikäyrä ja siihen sovitettu käyrä. Selkeyden vuoksi sovitettua
käyrää on siirretty ylöspäin.}
\label{theorpict}
\end{figure}

Sovitusalgoritmi konvergoitui 150 iteraation jälkeen, joten konvergenssiin
täytyi kokeilla $150\cdot150 = 22 500$ erilaista mahdollista ratkaisua. Algoritmin
löytämät kerrospaksuudet ja koostumukset ovat taulukossa \ref{theorfit}.
Tulevan säteen intensiteetille algoritmi löysi arvon 69,9 dB, taustasäteilyn intensiteetille
arvon $-0{,}9$ dB ja kulmalle $\phi$ arvon 0,0301\grad. Sovitusalgoritmin löytämien
arvojen poikkeama teoreettisen käyrän laskennassa käytetyistä arvoista selittyy sillä,
että todelliseen mittaukseen liittyvää epäideaalisuutta mallinnettiin kohinalla ja konvoluutiolla,
mikä siirsi optimia pois alunperin käytetyistä arvoista. Differentiaalievoluutioalgoritmi
siis selvisi tästä vaikeasta sovitusongelmasta ja ohjelma on siten käyttökelpoinen myös monimutkaisten
kerrosrakenteiden röntgendiffraktiomittauksiin. Kuvassa \ref{theorpict} on esitetty
alkuperäinen teoreettinen kohinainen käyrä ja siihen sovitettu käyrä.

\begin{table}[p!]
\caption{Sovitusalgoritmin testauksessa käytetty teoreettinen kerrosmalli ja
siihen liittyvät kerrosten ominaisuudet ja niiden sovitusvälit.}
\label{theorfit}
\begin{center}
\setlength{\tabcolsep}{1mm}
\setlength{\extracolsep}{1mm}
\begin{tabular}{|l|r|r|r|r|r|r|} \hline
Kerros & oikea   & oikea koo- & paksuuden   & koostumuksen & sovitettu & sovitettu \\
       & paksuus & stumus     & sovitusväli & sovitusväli  & paksuus   & koostumus \\ \hline
GaAs   & 22 nm   &            & 10--30 nm   &              & 22,0 nm   &           \\ \hline
InGaAs & 12 nm   & 0,12       & 10--30 nm   & 0,05--0,15   & 12,0 nm   & 0,119     \\ \hline
GaAs   & 18 nm   &            & 10--30 nm   &              & 18,0 nm   &           \\ \hline
InGaAs &  8 nm   & 0,08       & 10--30 nm   & 0,05--0,15   & 7,9 nm    & 0,081     \\ \hline
GaAs   & 20 nm   &            & 10--30 nm   &              & 20,0 nm   &           \\ \hline
InGaAs & 10 nm   & 0,10       & 10--30 nm   & 0,05--0,15   & 9,9 nm    & 0,100     \\ \hline
GaAs   & $\infty$&            & $\infty$    &              & $\infty$  &           \\ \hline
\end{tabular}
\end{center}
\end{table}

\clearpage
\section{Yhteenveto}

Tässä työssä käsiteltiin epitaktisten kerrosrakenteiden röntgendiffraktiota.
Aluksi esitettiin teoria, jonka avulla voidaan laskea diffraktiointensiteetti.
Intensiteettikäyrän simulointiin tarvitaan kiteen suskeptiivisuus, joka voidaan
esittää Fourier-sarjan avulla. Epitaktinen kerros venyy, ja venymät voidaan
laskea hilavakioista ja elastiikan laeista. Moniyhdistepuolijohteiden
hilavakiota ja Poissonin lukua voidaan approksimoida lineaarisella lailla, jota
hilavakion tapauksessa kutsutaan Vegardin laiksi. Fourier-sarjan kertoimet
voidaan laskea, kun tunnetaan atomien paikat, niiden elektronitiheyden
Fourier-muunnokset ja Hönlin korjaustermit, jotka kuvaavat aineen sähköisen
käyttäytymisen poikkeamaa yksinkertaisesta fysikaalisesta mallista eli
anomaalista dispersiota. Myös lämpövärähtelyt on otettava huomioon
Fourier-sarjan laskennassa Debye-Waller-kertoimella. 

Röntgendiffraktiota käsiteltiin dynaamisella diffraktioteorialla, joka perustuu
Maxwellin yhtälöihin. Siinä aine oletetaan ei-magneettiseksi ja homogeeniseksi,
eikä aineessa ole vapaata virrantiheyttä tai varaustiheyttä. Sähköinen
suskeptiivisuus esitetään Fourier-sarjan avulla. Monikerrosrakenteiden
diffraktiointensiteetti lasketaan näihin oletuksiin ja Maxwellin yhtälöihin
perustuvan Takagi-Taupin-differentiaaliyhtälön iteraatiokaavamuotoisesta
ratkaisusta.

Röntgendiffraktioyhtälöiden monimutkaisuudesta johtuen mittausdatasta ei saada
määritettyä kerrosten ominaisuuksia suoraan algebrallisesti tai
yksinkertaisella numeerisella yhtälönratkaisulla. Teoreettisen diffraktiokäyrän
simulointi on kuitenkin helppoa, jos tiedetään kerrosten ominaisuudet. Tämä
käänteisongelma ratkaistiin sovittamalla mittaustuloksiin käyrä
differentiaalievoluutioalgoritmilla, jonka risteytysoperaattoria parannettiin
pääkomponenttianalyysilla.

Tässä työssä kirjoitettiin Javalla ja Matlabilla graafinen tietokoneohjelma,
joka osaa simuloida röntgendiffraktiokäyriä ja sovittaa simuloidun käyrän
mittaukseen. Ohjelmalla voidaan määrittää monikerrosrakenteen kerrosten paksuus
ja koostumus tai relaksaatio. Ohjelma käyttöliittymä pyrittiin sellaiseksi,
että sillä pystytään muokkaamaan kerrosmalleja mahdollisimman tehokkaasti.
Ohjelman havaittiin simuloivan samanlaisia röntgendiffraktiokäyriä kuin muutkin
ohjelmat, joten ohjelman kriittisin osa eli diffraktiokäyrien laskenta on
kunnossa. Uudella ohjelmalla analysoitiin neljä aiemmin tehtyä mittausta, ja se
antoi samanlaiset tulokset kuin Philipsin ohjelma. Lisäksi sovitusalgoritmin
havaittiin toimivan monimutkaiselle monikerrosrakenteelle 12
sovitusparametrilla.

Vaikka tietokoneohjelma tekee samat asiat kuin kaupalliset ohjelmat, sillä on
huomattavasti etuja näihin nähden. Ohjelma toimii täysin ilman mitään
kaupallisia komponentteja. Käyttöliittymä on erittäin tehokas, mikä tekee
kerrosmallin rakentamisesta ja muokkaamisesta nopeaa. Lähdekoodi on saatavilla,
joten ohjelmaa voidaan muuttaa tarpeita vastaavaksi. Ohjelman sovitusalgoritmi
toimii monimutkaisillekin monikerrosrakenteille.

Vaikka röntgenheijastusohjelma saatiin toimimaan, siinä on vielä huomattavasti
parantamisen varaa. Sekä ohjelmakoodin että sovitusalgoritmin
toimintaperiaatteen voisi ohjelmoida huomattavasti tehokkaammaksi.
Ohjelmointikielen vaihtaminen esimerkiksi C++:ksi auttaisi tässä. Koska
ohjelman lähdekoodi on saatavilla, sitä voidaan käyttää varsinaisten mittausten
lisäksi itse mittausmenetelmään liittyvään tutkimukseen. Mikro- ja
nanotekniikan laboratoriossa on tutkittu kohinan vaikutusta
röntgenheijastusmittausten tarkkuuteen\cite{accuracy}. Luvussa \ref{meas}
havaittiin, että simuloitu kohina vääristi sovituksesta saatuja tuloksia.
Ohjelman lähdekoodia voidaan käyttää apuna tarkemmassa tutkimuksessa
röntgendiffraktiomittausten tarkkuudesta.

\clearpage
\addcontentsline{toc}{section}{Viitteet}

\bibliographystyle{kandi}
\bibliography{kandi}

\end{document}